\documentclass{article}
\usepackage{graphicx}
\usepackage{amsmath}
\usepackage{amssymb}

\begin{document}

\title{\textbf{Lagrangian and orthogonal splittings, quasitriangular Lie
bialgebras and almost complex product structures}\\
\smallskip\ }
\author{\textbf{H. Montani{\thanks{%
e-mail: \textit{hmontani@uaco.unpa.edu.ar }} }} \\
Departamento de Ciencias Exactas y Naturales\\
Unidad Acad\'{e}mica Caleta Olivia,\\
Universidad Nacional de la Patagonia Austral \\
and\\
CIT Golfo de San Jorge - CONICET,\\
(9011) Caleta Olivia, Argentina.}
\maketitle

\begin{abstract}
We study Lagrangian and orthogonal splittings\textbf{\ }of quadratic vector
spaces establishing an equivalence with complex product structures. We show
that a Manin triple equipped with generalized metric $\mathcal{G}+\mathcal{B}
$ such that $\mathcal{B}$ is an $\mathcal{O}$-operator with extension $%
\mathcal{G}$ of mass -1 can be turned into another Manin triple that admits
also an orthogonal splitting in\textbf{\ }Lie ideals. Conversely, a
quadratic Lie algebra orthogonal direct sum of a pair anti-isomorphic Lie
algebras, following similar steps as in the previous case, can be turned
into a Manin triple admitting an orthogonal splitting into Lie ideals.
\end{abstract}

\tableofcontents

\bigskip

\section{\label{sec:level1}Introduction}

This work aims to study some algebraic aspects of quadratic vector spaces
that admit a Lagrangian (maximally isotropic) and an orthogonal direct sum
decomposition associated with complex product structures, mainly when they
are endowed with a Lie algebra structure. In particular, we study a Manin
triple with an orthogonal vector space splitting and use the (almost)
complex product structure to build a new Manin triple where the orthogonal
subspaces become Lie ideals. Conversely, starting from a pair of
anti-isomorphic Lie algebras that are orthogonal relative to a
non-degenerate symmetric bilinear form on the Lie algebra direct sum, a
Manin triple is obtained that has these Lie algebras as Lie ideals. All of
this is done by resorting to some tools in the realm of the modified
classical Yang-Baxter equation.

Lagrangian and orthogonal splitting of vector spaces intervene as important
ingredients in the formulation of Poisson-Lie T-duality: the Lagrangian
decomposition is inherent to the Manin triple on which the T-dual sigma
models are built, while projections on orthogonal subspaces give rise to the
relevant dynamics \cite{Klim-Sev-00},\cite{Klim-Sev-01},\cite{Klimcik-01},%
\cite{Klim-Sev-02}. This orthogonal splitting, in subspaces of the same
dimension, can be regarded as the eigenspaces decomposition of an involutive
operator $\mathcal{E}$ that encodes the information of a generalized metric.
Also, a linear operator like $\mathcal{E}$ appears in string theory
T-duality, through Double Field Theory approach, disguised by a right
multiplication with a null signature metric analogous to the bilinear form
provided by the Manin triple (see for instance the review in ref. \cite{AMN}%
).

Hence we study the coexistence of Lagrangian and orthogonal splitting in
quadratic vector spaces and in Manin triples, when they arise from (almost)
complex product structures \cite{Andrada-Salamon} composed of a product
structure $\mathcal{E}$ and an almost complex structure $\mathcal{J}$
naturally associated with $\mathcal{E}$. These operators are tied to a
generalized metric on a Lagrangian splitting, or an anti-isomorphism between
the subspaces of an orthogonal splitting. From a given pair $\left\{ 
\mathcal{E},\mathcal{J}\right\} $, a family of Lagrangian or orthogonal
splittings can be obtained by gauge transformations \cite{Severa-Weinstein}.
Then, in the framework of Lie bialgebras, appealing to the formulation by $%
\mathcal{O}$-operators \cite{Chengming-01} of the modified classical
Yang-Baxter equation, we use quasitriangular factorizable solutions to get a
Manin triple where the orthogonal subspaces become Lie subalgebras, Lie
ideals in fact, and to the twilled extension procedure \cite%
{Kosmann-Magri-01} to assemble the Lagrangian components in a bigger Lie
algebra. The construction can be reversed, allowing the construction of a
Manin triple with the same properties out of a pair of anti-isomorphic Lie
algebras. Again, in this procedure, quasitriangular factorizable solutions
of the modified classical Yang-Baxter equation are used.

We carry out this work in the following steps. In Section 2, we consider a $%
2n$-dimensional real quadratic vector\textit{\ }space with a split bilinear
form\textit{\ }$\left( ,\right) _{\mathrm{V}}:\mathrm{V}\otimes \mathrm{V}%
\longrightarrow \mathbb{R}$ endowed with a symmetric complex product
structure $\left\{ \mathcal{E},\mathcal{J}\right\} $ and study the direct
sum decompositions (\emph{splittings}) $\mathrm{V}=\mathrm{E}^{+}\oplus 
\mathrm{E}^{-}=\mathrm{F}_{+}\oplus \mathrm{F}_{-}$ with $\mathrm{E}^{\pm }$
an $n$-dimensional orthogonal subspace, and $\mathrm{F}_{\pm }$ a Lagrangian
(maximally isotropic) subspace, arising from the involutive operators $%
\mathcal{E}$ and $\mathcal{JE}$. In Section 3, we briefly describe the
general framework to deal with Lie algebras and complex product structures.
In Section 4, we start with a Manin triple $\left( \mathfrak{g},\mathfrak{g}%
_{+},\mathfrak{g}_{-}\right) $ endowed with a generalized metric $\mathcal{G}%
+\mathcal{B}:\mathfrak{g}_{+}\longrightarrow \mathfrak{g}_{-}$, which gives
rise to an orthogonal splitting $\mathfrak{g}=\mathcal{E}^{+}\oplus \mathcal{%
E}^{-}$, and we promote it to an $\mathcal{O}$-operator to get a new Lie
algebra structure on $\mathfrak{g}_{+}$ by using a quasitriangular
factorizable solution of the modified classical Yang-Baxter equation. Then
we built a new Manin triple as a twilled extension of these Lie subalgebras,
which has the notorious property of admitting the orthogonal subspaces as
Lie ideals. In Section 5, we start with a pair of anti-isomorphic $n$%
-dimensional Lie algebras $\mathrm{E}^{+}$ and $\mathrm{E}^{-}$ and joint
them together in a quadratic Lie algebra direct sum $\mathfrak{g}$, such
that $\mathrm{E}^{+}$ and $\mathrm{E}^{-}$ are mutually orthogonal. From the
associated operator $\mathcal{J}$, we construct a Lagrangian splitting $%
\mathfrak{g}=\mathrm{F}_{+}\oplus \mathrm{F}_{-}$. Again, from a metric $%
\mathcal{G}$ on $\mathrm{F}_{+}$ and a gauge transformation by a
skew-symmetric map $\mathcal{B}$ we get a new orthogonal splitting of $%
\mathfrak{g}$. Following the analogous steps as in the previous section, we
finally get a Manin triple admitting the orthogonal splitting as Lie ideals.
Finally, in Section 6, some conclusion are summarized.


\section{\label{cp vector spaces}Complex product structure on vector spaces:
orthogonal and Lagrangian splittings}

Let us review some basic facts about real quadratic vector spaces (see for
instance ref. \cite{Meinrenken-book}). A \emph{quadratic vector space} $%
\mathrm{V}$ is a vector space endowed with non-degenerate symmetric bilinear
form $\left( ,\right) _{\mathrm{V}}:\mathrm{V}\otimes \mathrm{V}%
\longrightarrow \mathbb{R}$. Given a subspace $\mathrm{U}\subset \mathrm{V}$%
, $\mathrm{U}^{\bot }$ denotes the orthogonal complement of $\mathrm{U}$. A
subspace $\mathrm{U}\subset \mathrm{V}$ is isotropic if $\mathrm{U}\subseteq 
\mathrm{U}^{\bot }$. If $\mathrm{V}$ is even dimensional, the bilinear form
is called \emph{split} if the maximally isotropic subspaces are of dimension 
$\frac{1}{2}\dim \mathrm{V}$, in this case these subspaces are named \emph{%
Lagrangian} subspaces.

We study the simultaneous decomposition of a quadratic vector space in a
direct sum of two Lagrangian and a direct sum of two orthogonal subspaces of
the same dimension. Let $\mathrm{V}$ be a $2n$-dimensional real quadratic
vector\textit{\ }space with a split bilinear form\textit{\ }$\left( ,\right)
_{\mathrm{V}}:\mathrm{V}\otimes \mathrm{V}\longrightarrow \mathbb{R}$, and
consider the direct sum decompositions $\mathrm{V}=\mathrm{E}^{+}\oplus 
\mathrm{E}^{-}=\mathrm{F}_{+}\oplus \mathrm{F}_{-}$ such that $\left\{ 
\mathrm{E}^{+},\mathrm{E}^{-}\right\} $ are mutually orthogonal $n$%
-dimensional subspaces while $\left\{ \mathrm{F}_{+},\mathrm{F}_{-}\right\} $
are Lagrangian subspaces. For the sake of brevity, we refer to this pair of
simultaneous direct sum decompositions of $\mathrm{V}$ as a \emph{double
splitting}.

A \emph{complex\ product\ structure} on the vector space $\mathrm{V}$ (see
ref. \cite{Andrada-Salamon}) is a pair of linear operators $\left\{ \mathcal{%
E},\mathcal{J}\right\} :\mathrm{V}\longrightarrow \mathrm{V}$ such that 
\begin{equation}
\begin{array}{lllll}
\mathcal{E}^{2}=\mathcal{I} & , & \mathcal{J}^{2}=-\mathcal{I} & , & 
\mathcal{EJ}+\mathcal{JE}=0%
\end{array}
\label{cpd 00a}
\end{equation}%
where $\mathcal{I}$ is the identity on $\mathrm{V}$. The involutive operator 
$\mathcal{E}$ is called \emph{product structure }on\emph{\ }the vector space 
$\mathrm{V}$ and the linear operator $\mathcal{J}$ is a complex structure on 
$\mathrm{V}$. Since $\mathcal{E}$ and $\mathcal{J}$ anticommute, $\mathcal{J}
$ is a linear bijection between the eigenspaces $\mathcal{E}^{\pm }$,
associated with the eigenvalues $\pm 1$ of $\mathcal{E}$, so the eigenspaces 
$\mathcal{E}^{\pm }$ are $n$-dimensional and $\mathcal{E}$ is called a \emph{%
paracomplex structures} \cite{Liberman}. There is a third linear operator,
namely $\mathcal{F}=\mathcal{JE}$, which is involutive and anticommutes with 
$\mathcal{E}$ and $\mathcal{J}$. Thus $\left\{ \mathcal{E},\mathcal{J},%
\mathcal{F}\right\} $ spans the algebra of linear homogeneous polynomials on
these operators.

Let $\mathrm{V}$ be a real quadratic vector\textit{\ }space, then we call $%
\left\{ \mathcal{E},\mathcal{J}\right\} $ a \emph{symmetric complex product
structure }when $\mathcal{E}$ and $\mathcal{J}$ are symmetric relative to
the bilinear form $\left( ,\right) _{\mathrm{V}}$, implying that $\mathcal{F}
$ is skew-symmetric. In this case, $\mathcal{J}$ is anti-compatible with $%
\left( ,\right) _{\mathrm{V}}$ so $\left( \left( ,\right) _{\mathrm{V}},%
\mathcal{J}\right) $ is an anti-Hermitian structure on $\mathrm{V}$.
However, from the complex product structure we may define a second quadratic
bilinear form on $\mathrm{V}$, namely $\left( ,\right) _{\mathcal{E}}:%
\mathrm{V}\otimes \mathrm{V}\longrightarrow \mathbb{R}$ as $\left(
X,Y\right) _{\mathcal{E}}:=\left( X,\mathcal{E}Y\right) _{\mathrm{V}}$ so
that $\left( \left( ,\right) _{\mathrm{V}},\mathcal{J}\right) $ is an
Hermitian structure.

A symmetric complex product structure $\left\{ \mathcal{E},\mathcal{J}%
\right\} $ on a quadratic vector space $\mathrm{V}$ has associated a double
splitting $\mathrm{V}=\mathrm{E}^{+}\oplus \mathrm{E}^{-}=\mathrm{F}%
_{+}\oplus \mathrm{F}_{-}$ defined by the eigenspaces $\mathrm{E}^{+}$ and $%
\mathrm{E}^{-}$ of the involutive symmetric operator $\mathcal{E}$ and the
eigenspaces $\mathrm{F}_{+}$ and $\mathrm{F}_{-}$ of the involutive
skew-symmetric operator $\mathcal{F}=\mathcal{JE}$. This implies the
existence of a linear bijections $\varphi :\mathrm{E}^{+}\longrightarrow 
\mathrm{E}^{-}$, with $\varphi ^{\top }=-\varphi ^{-1}$, such that the
matrix block form of the operators $\left\{ \mathcal{E},\mathcal{J},\mathcal{%
F}\right\} $ in a basis of eigenvectors of $\mathcal{E}$ are%
\begin{equation}
\begin{array}{cc}
\mathcal{E} & =\left( 
\begin{array}{cc}
I & 0 \\ 
0 & -I%
\end{array}%
\right) \,\,\,\,,\,\,\,\,\mathcal{F}=\left( 
\begin{array}{cc}
0 & \varphi ^{-1} \\ 
\varphi & 0%
\end{array}%
\right) \,\,\,\,,\,\,\,\,\mathcal{J}=\mathcal{FE}=\left( 
\begin{array}{cc}
0 & -\varphi ^{-1} \\ 
\varphi & 0%
\end{array}%
\right)%
\end{array}%
,  \label{cpd 05b}
\end{equation}%
and of a linear bijection $\mathcal{G}:\mathrm{F}_{+}\longrightarrow \mathrm{%
F}_{-}$, with $\mathcal{G}^{\top }=\mathcal{G}$, such that the matrix block
form of the operators $\left\{ \mathcal{E},\mathcal{J},\mathcal{F}\right\} $
in a basis of eigenvectors of $\mathcal{F}$ are 
\begin{equation*}
\begin{array}{cc}
\mathcal{E} & =\left( 
\begin{array}{cc}
0 & \mathcal{G}^{-1} \\ 
\mathcal{G} & 0%
\end{array}%
\right) \,\,\,\,,\,\,\,\,\mathcal{F}=\left( 
\begin{array}{cc}
I & 0 \\ 
0 & -I%
\end{array}%
\right) \,\,\,\,,\,\,\,\,\mathcal{J}=\mathcal{FE}=\left( 
\begin{array}{cc}
0 & \mathcal{G}^{-1} \\ 
-\mathcal{G} & 0%
\end{array}%
\right)%
\end{array}%
.
\end{equation*}%
In the first case the eigenspace $\mathrm{F}_{\pm }$ can be identified as
the graph of $\varphi $ 
\begin{equation*}
\mathrm{F}_{\pm }=\left\{ X^{+}\pm \varphi \left( X^{+}\right) /X^{+}\in 
\mathrm{E}^{+}\right\} =\mathrm{graph}\left( \pm \varphi \right)
\end{equation*}%
and the linear map $\mathcal{G}:\mathrm{F}_{+}\longrightarrow \mathrm{F}_{-}$
is realized as 
\begin{equation*}
\mathcal{G}\left( X^{+}+\varphi \left( X^{+}\right) \right) =X^{+}-\varphi
\left( X^{+}\right) .
\end{equation*}%
In the second case, the eigenspace $\mathrm{E}^{\pm }$ coincides with the
graph of $\mathcal{G}$ 
\begin{equation*}
\mathrm{E}^{\pm }=\left\{ X_{+}\pm \mathcal{G}\left( X_{+}\right) /X_{+}\in 
\mathrm{F}_{+}\right\} =\mathrm{graph}\left( \pm \mathcal{G}\right) .
\end{equation*}%
and the linear map $\varphi :\mathrm{E}^{+}\longrightarrow \mathrm{E}^{-}$
turn to be 
\begin{equation}
\varphi \left( X_{+}+\mathcal{G}\left( X_{+}\right) \right) =X_{+}-\mathcal{G%
}\left( X_{+}\right)  \label{cpd 05a}
\end{equation}

Thus, double splittings induced by complex product structures can be
equivalently obtained from an orthogonal splitting supplied with a linear
bijection $\varphi :\mathrm{E}^{+}\longrightarrow \mathrm{E}^{-}$ with $%
\varphi ^{\top }=-\varphi ^{-1}$ or, alternatively, from a Lagrangian
splitting supplied with a linear bijection $\mathcal{G}:\mathrm{F}%
_{+}\longrightarrow \mathrm{F}_{-}$ with $\mathcal{G}^{\top }=\mathcal{G}$.
Note that $\mathcal{G}$ can be regarded as a \emph{metric} on the Lagrangian
component $\mathrm{F}_{+}$.

In the next sections we shall exploit these alternative ways of building
double splittings from a complex product structure to obtain a direct sum
Lie algebra from an special Manin triple and, reciprocally, to obtain a
Manin triple from a particular Lie algebra direct sum.

Starting from a Lagrangian splitting $\mathrm{V}=\mathrm{F}_{+}\oplus 
\mathrm{F}_{-}$ and a linear map $\mathcal{G}:\mathrm{F}_{+}\longrightarrow 
\mathrm{F}_{-}$ we can obtain a wide family of orthogonal splittings through
a class of isometries named \emph{gauge transformations }\cite%
{Severa-Weinstein}. A \emph{twisting }or\emph{\ gauge transformation }on the
vector space $\mathrm{V}$ is implemented on the Lagrangian decomposition $%
\mathrm{V}=\mathrm{F}_{+}\oplus \mathrm{F}_{-}$ by a skew-symmetric linear
map $\mathcal{B}:\mathrm{F}_{+}\longrightarrow \mathrm{F}_{-}$ such that 
\begin{equation}
\mathcal{B}\cdot \left( X_{+}+X_{-}\right) =X_{+}+X_{-}+\mathcal{B}\left(
X_{+}\right) .  \label{cpd 06c}
\end{equation}%
So, we get the family of subspaces parametrized by $\mathcal{B}\in \mathrm{%
Skew}\left( \mathrm{F}_{+},\mathrm{F}_{-}\right) $ 
\begin{equation}
\mathrm{E}_{\mathcal{B}}^{\pm }=\left\{ X_{+}\pm \left( \mathcal{B}\pm 
\mathcal{G}\right) X_{+}/X_{+}\in \mathrm{F}_{+}\right\} =\mathrm{graph}%
\left( \mathcal{B}\pm \mathcal{G}\right) ,  \label{cpd 06d}
\end{equation}%
such that $\mathrm{V}=\mathrm{E}_{\mathcal{B}}^{+}\oplus \mathrm{E}_{%
\mathcal{B}}^{-}$ is an orthogonal splitting of $\mathrm{V}$. Note that if $%
\mathcal{G}$ is positive definite, $\mathcal{G}\pm \mathcal{B}$ is
invertible and can be regarded as a generalized metric on $\mathrm{F}_{+}$.
However, we will abuse the language and keep this name even in the case
where $\mathcal{G}\pm \mathcal{B}$ is non-invertible.

Gauge transformations led us to a broader framework in which the vector
space $\mathrm{V}$ admit simultaneously the Lagrangian splitting $\mathrm{V}=%
\mathrm{F}_{+}\oplus \mathrm{F}_{-}$ and the orthogonal one $\mathrm{V}=%
\mathrm{E}_{\mathcal{B}}^{+}\oplus \mathrm{E}_{\mathcal{B}}^{-}$, for
arbitrary skew-symmetric linear map $\mathcal{B}:\mathrm{F}%
_{+}\longrightarrow \mathrm{F}_{-}$. So we must review the above block
matrix realization of the associated complex product structure $\left\{ 
\mathcal{E_{\mathcal{B}}},\mathcal{J_{\mathcal{B}}}\right\} $.

\begin{description}
\item[Proposition] \textit{The complex product structure} $\left\{ \mathcal{%
E_{\mathcal{B}}},\mathcal{J_{\mathcal{B}}}\right\} $ \textit{associated with
the decompositions} $\mathrm{V}=\mathrm{E}_{\mathcal{B}}^{+}\oplus \mathrm{E}%
_{\mathcal{B}}^{-}=\mathrm{F}_{+}\oplus \mathrm{F}_{-}$ \textit{is
represented in the Lagrangian splitting as} 
\begin{equation}
\begin{array}{ccc}
\mathcal{E}_{\mathcal{B}}=\left( 
\begin{array}{cc}
-\mathcal{G}^{-1}\mathcal{B} & \mathcal{G}^{-1} \\ 
\mathcal{G}-\mathcal{BG}^{-1}\mathcal{B} & \mathcal{BG}^{-1}%
\end{array}%
\right) & , & \mathcal{J}_{\mathcal{B}}=\left( 
\begin{array}{cc}
-\mathcal{G}^{-1}\mathcal{B} & \mathcal{G}^{-1} \\ 
-\mathcal{G}-\mathcal{BG}^{-1}\mathcal{B} & \mathcal{BG}^{-1}%
\end{array}%
\right)%
\end{array}
\label{cpd 07a}
\end{equation}%
\textit{while in the orthogonal splitting is represented as} 
\begin{equation*}
\begin{array}{ccc}
\mathcal{E_{\mathcal{B}}}=\left( 
\begin{array}{cc}
I & 0 \\ 
0 & -I%
\end{array}%
\right) & , & \mathcal{J_{\mathcal{B}}}=\left( 
\begin{array}{cc}
0 & -\varphi _{\mathcal{B}}^{-1} \\ 
\varphi \mathcal{_{\mathcal{B}}} & 0%
\end{array}%
\right)%
\end{array}%
\end{equation*}%
\textit{where} $\varphi \mathcal{_{\mathcal{B}}}:\mathcal{E}_{\mathcal{B}%
}^{+}\longrightarrow \mathcal{E}_{\mathcal{B}}^{-}$ \textit{defined as} 
\begin{equation}
\varphi \mathcal{_{\mathcal{B}}}\left( X_{+}+\left( \mathcal{B}+\mathcal{G}%
\right) X_{+}\right) =X_{+}+\left( \mathcal{B}-\mathcal{G}\right) X_{+}.
\label{cpd 07b}
\end{equation}
\end{description}

\textbf{Proof: }The proof is straightforward, it is easy to check that $%
\left( \mathcal{E}_{\mathcal{B}}\right) ^{2}=-\left( \mathcal{J}_{\mathcal{B}%
}\right) ^{2}=\mathcal{I}$, and $\mathcal{E}_{\mathcal{B}}\mathcal{J}_{%
\mathcal{B}}+\mathcal{J}_{\mathcal{B}}\mathcal{E}_{\mathcal{B}}=I$, and that
the eigenspaces of $\mathcal{E}_{\mathcal{B}}$ are $\mathrm{E}_{\mathcal{B}%
}^{+}$ and $\mathrm{E}_{\mathcal{B}}^{-}$ in both direct sum decompositions. 
$\blacksquare $

The operators $\mathcal{E_{\mathcal{B}}}$ and $\mathcal{J_{\mathcal{B}}}$
introduced in eq. $\left( \ref{cpd 07a}\right) $ are uniquely defined up to
a conformal factor steaming from the metric change $\mathcal{G}
\longrightarrow e ^{\phi}\mathcal{G}$.

The operator $\mathcal{F}_{\mathcal{B}}$ in $\mathrm{V}=\mathrm{F}_{+}\oplus 
\mathrm{F}_{-}$ is 
\begin{equation*}
\mathcal{F}_{\mathcal{B}}=\left( 
\begin{array}{cc}
I & 0 \\ 
2\mathcal{B} & -I%
\end{array}%
\right)
\end{equation*}%
with the eigenspaces $\mathrm{F}_{\mathcal{B}-}=\mathrm{F}_{-}$ and $\mathrm{%
F}_{\mathcal{B}+}=\mathcal{B}\cdot \mathrm{F}_{+}$, providing a family of
Lagrangian splittings parametrized by skew-symmetric maps from $\mathrm{F}%
_{+}\ $to $\mathrm{F}_{-}$.

The operators $\left( \ref{cpd 07a}\right) $ can be written as \textit{\ }%
\begin{equation*}
\begin{array}{ccc}
\mathcal{E}=\mathrm{B}^{\top }\mathrm{GBA} & , & \mathcal{J}=\mathrm{B}%
^{\top }\mathrm{JBA}%
\end{array}%
\end{equation*}%
where%
\begin{equation*}
\begin{array}{ccc}
\mathrm{B}=\left( 
\begin{array}{cc}
\mathcal{I} & -\mathcal{B} \\ 
0 & -\mathcal{I}%
\end{array}%
\right) & , & \mathrm{G}=\left( 
\begin{array}{cc}
\mathcal{G}^{-1} & 0 \\ 
0 & \mathcal{G}%
\end{array}%
\right) , \\ 
&  &  \\ 
\mathrm{J}=\left( 
\begin{array}{cc}
\mathcal{G}^{-1} & 0 \\ 
0 & -\mathcal{G}%
\end{array}%
\right) & , & \mathrm{A}=\left( 
\begin{array}{cc}
0 & \mathcal{I} \\ 
\mathcal{I} & 0%
\end{array}%
\right) .%
\end{array}%
\end{equation*}

It is interesting to note that linear operators like to $\mathcal{E}$,
disguised as 
\begin{equation*}
\begin{array}{c}
\left( 
\begin{array}{cc}
\mathcal{G}^{-1} & -\mathcal{G}^{-1}\mathcal{B} \\ 
\mathcal{BG}^{-1} & \mathcal{G}-\mathcal{BG}^{-1}\mathcal{B}%
\end{array}%
\right) =\left( 
\begin{array}{cc}
-\mathcal{G}^{-1}\mathcal{B} & \mathcal{G}^{-1} \\ 
\mathcal{G}-\mathcal{BG}^{-1}\mathcal{B} & \mathcal{BG}^{-1}%
\end{array}%
\right) \left( 
\begin{array}{cc}
0 & I \\ 
I & 0%
\end{array}%
\right) ,%
\end{array}%
{}
\end{equation*}%
is pervasive in string theory T-duality. There its occurrence can be traced
back to references \cite{Scherk-Schwarz-01}, \cite{Maharana-Schwarz-01}
where it appears as a \emph{metric} on a $2D$-dimensional manifold, then in
early studies on T-duality as a \emph{generalized metric} and, more
recently, it becomes in the central object in the so called Double Field
Theory approach to T-duality \cite{Hull-Zwiebach-00},\cite%
{Hohm-Hull-Zwiebach-01}. It also plays a relevant role in generalized K\"{a}%
hler geometry \cite{Gualtieri-Th}. Hence, all these problems can be endowed
with an (almost) complex structure $\left\{ \mathcal{E},\mathcal{J}\right\} $
like that given in eq. $\left( \ref{cpd 07a}\right) $.

\subsection{Dual construction}

An analogue of the above construction can be achieved using the inverse map $%
\varphi ^{-1}:\mathrm{E}^{-}\longrightarrow \mathrm{E}^{+}$ in place of $%
\varphi :\mathrm{E}^{+}\longrightarrow \mathrm{E}^{-}$, in such a way that
the eigenspaces of $\mathrm{F}$ are now defined as 
\begin{equation*}
\mathrm{F}_{\pm }=\left\{ X^{-}\pm \varphi ^{-1}\left( X^{-}\right)
/X^{-}\in \mathrm{E}^{-}\right\}
\end{equation*}%
This construction comes to be dual of the former, relative to the bilinear
form $\left( ,\right) _{\mathrm{V}}$ in the following sense%
\begin{equation*}
\left( X^{-}\pm \varphi ^{-1}\left( X^{-}\right) ,Y^{+}\right) _{\mathrm{V}%
}=\left( X^{-},Y^{+}\mp \varphi \left( Y^{+}\right) \right) _{\mathrm{V}},
\end{equation*}%
for $X^{-}\in \mathrm{E}^{-},Y^{+}\in \mathrm{E}^{+}$. The analogous of the
map $\mathcal{G}$ is now the linear bijection $\mathcal{\tilde{G}}:\mathrm{F}%
_{-}\longrightarrow \mathrm{F}_{+}$ defined as%
\begin{equation*}
\mathcal{\tilde{G}}\left( X^{-}-\varphi ^{-1}\left( X^{-}\right) \right)
=X^{-}+\varphi ^{-1}\left( X^{-}\right)
\end{equation*}%
and from it we get 
\begin{equation*}
\mathrm{E}^{\mp }=\left\{ X_{-}\pm \mathcal{\tilde{G}}\left( X_{-}\right)
/X_{-}\in \mathrm{F}_{-}\right\} =\mathrm{graph}\left( \pm \mathcal{\tilde{G}%
}\right) .
\end{equation*}%
By writing $X^{-}=-\varphi \left( X^{+}\right) $ one may see that $\mathcal{%
\tilde{G}}=\mathcal{G}^{-1}$. In the Lagrangian decomposition $\mathrm{V}=%
\mathrm{F}_{+}\oplus \mathrm{F}_{-}$ the operators $\mathcal{\tilde{E}}$ and 
$\mathcal{\tilde{J}}$ are represented by the block matrices%
\begin{equation*}
\begin{array}{ccc}
\mathcal{\tilde{E}}=\left( 
\begin{array}{cc}
0 & \mathcal{\tilde{G}} \\ 
\mathcal{\tilde{G}}^{-1} & 0%
\end{array}%
\right) & , & \mathcal{\tilde{J}}=\left( 
\begin{array}{cc}
0 & \mathcal{\tilde{G}} \\ 
-\mathcal{\tilde{G}}^{-1} & 0%
\end{array}%
\right)%
\end{array}%
.
\end{equation*}

Things become more interesting after applying the gauge transformation\emph{%
\ }isometry 
\begin{equation*}
\mathcal{\tilde{B}}\cdot \left( X_{+}+X_{-}\right) =\left( X_{+}+\mathcal{%
\tilde{B}}\left( X_{-}\right) +X_{-}\right)
\end{equation*}%
with $\mathcal{\tilde{B}}:\mathrm{F}_{-}\longrightarrow \mathrm{F}_{+}$ a
skew-symmetric linear map, which allows to get the family of orthogonal
subspaces 
\begin{equation*}
\mathrm{E}_{\mathcal{\tilde{B}}}^{\pm }=\left\{ X_{-}\pm \left( \mathcal{%
\tilde{B}}\pm \mathcal{\tilde{G}}\right) X_{-}/X_{-}\in \mathrm{F}%
_{-}\right\} =\mathrm{graph}\left( \mathcal{\tilde{B}}\pm \mathcal{\tilde{G}}%
\right) ,
\end{equation*}%
parametrized by $\mathcal{\tilde{B}}\in \mathrm{Skew}\left( \mathrm{F}_{-},%
\mathrm{F}_{+}\right) $. Hence, the Lagrangian and orthogonal splitting $%
\mathrm{V}=\mathrm{F}_{+}\oplus \mathrm{F}_{-}=\mathrm{E}_{\mathcal{\tilde{B}%
}}^{+}\oplus \mathrm{E}_{\mathcal{\tilde{B}}}^{-}$ have associated the
complex product structure $\left\{ \mathcal{\tilde{E}_{\mathcal{B}}},%
\mathcal{\tilde{J}_{\mathcal{B}}}\right\} $ 
\begin{equation*}
\begin{array}{ccc}
\mathcal{E_{\mathcal{\tilde{B}}}}=\left( 
\begin{array}{cc}
I & 0 \\ 
0 & -I%
\end{array}%
\right) & , & \mathcal{J_{\mathcal{\tilde{B}}}}=\left( 
\begin{array}{cc}
0 & -\tilde{\varphi}^{-1} \\ 
\tilde{\varphi} & 0%
\end{array}%
\right)%
\end{array}%
,
\end{equation*}%
referred to $\mathrm{V}=\mathrm{E}_{\mathcal{\tilde{B}}}^{+}\oplus \mathrm{E}%
_{\mathcal{\tilde{B}}}^{-}$, where $\tilde{\varphi}:\mathcal{E}_{\mathcal{B}%
}^{+}\longrightarrow \mathcal{E}_{\mathcal{B}}^{-}$ now defined as 
\begin{equation*}
\tilde{\varphi}\left( X_{-}+\left( \mathcal{\tilde{B}}+\mathcal{\tilde{G}}%
\right) X_{-}\right) =X_{-}+\left( \mathcal{\tilde{B}}-\mathcal{G}\right)
X_{-}.
\end{equation*}%
In the Lagrangian splitting $\mathrm{V}=\mathrm{F}_{+}\oplus \mathrm{F}_{-}$
they are represented by block matrices \cite{Klimcik-01}%
\begin{equation*}
\begin{array}{ccc}
\mathcal{E}_{\mathcal{\tilde{B}}}=\left( 
\begin{array}{cc}
\mathcal{\tilde{B}\tilde{G}}^{-1} & \mathcal{\tilde{G}}-\mathcal{\tilde{B}%
\tilde{G}}^{-1}\mathcal{\tilde{B}} \\ 
\mathcal{\tilde{G}}^{-1} & -\mathcal{\tilde{G}}^{-1}\mathcal{\tilde{B}}%
\end{array}%
\right) & , & \mathcal{\tilde{J}}_{\mathcal{\tilde{B}}}=\left( 
\begin{array}{cc}
\mathcal{\tilde{B}\tilde{G}}^{-1} & -\mathcal{\tilde{G}}-\mathcal{\tilde{B}%
\tilde{G}}^{-1}\mathcal{\tilde{B}} \\ 
\mathcal{\tilde{G}}^{-1} & -\mathcal{\tilde{G}}^{-1}\mathcal{\tilde{B}}%
\end{array}%
\right) .%
\end{array}%
\end{equation*}

The eigenspace $\mathrm{E}_{\mathcal{\tilde{B}}}^{\pm }$ of $\mathcal{\tilde{%
E}}_{\mathcal{\tilde{B}}}$ coincides with $\mathrm{E}_{\mathcal{B}}^{\pm }$ $%
\left( \ref{cpd 06d}\right) $ if and only if%
\begin{equation*}
\begin{array}{ccc}
\left( \mathcal{B}\pm \mathcal{G}\right) \left( \mathcal{\tilde{B}}\pm 
\mathcal{\tilde{G}}\right) =\mathcal{I} & \text{and} & \left( \mathcal{%
\tilde{B}}\pm \mathcal{\tilde{G}}\right) \left( \mathcal{B}\pm \mathcal{G}%
\right) =\mathcal{I}%
\end{array}%
,
\end{equation*}%
which in turn implies the relations 
\begin{eqnarray*}
\mathcal{\tilde{G}} &=&\left( \mathcal{G}-\mathcal{BG}^{-1}\mathcal{B}%
\right) ^{-1} \\
&& \\
\mathcal{\tilde{B}} &=&-\mathcal{G}^{-1}\mathcal{B}\left( \mathcal{G}-%
\mathcal{BG}^{-1}\mathcal{B}\right) ^{-1}=-\left( \mathcal{G}-\mathcal{BG}%
^{-1}\mathcal{B}\right) ^{-1}\mathcal{BG}^{-1}
\end{eqnarray*}%
making $\mathcal{E_{\mathcal{\tilde{B}}}}=\mathcal{E_{\mathcal{B}}}$. These
relations also hold after interchanging $\left( \mathcal{G},\mathcal{B}%
\right) \longleftrightarrow \left( \mathcal{\tilde{G}},\mathcal{\tilde{B}}%
\right) $. Note that both descriptions lead to the same double splitting
provided the generalized metric $\mathcal{G}\pm \mathcal{B}$ provided it is
invertible which, for instance, is warranted if $\mathcal{G}$ gives rise to
a positive definite metric on $\mathrm{F}_{+}$, or $\mathcal{\tilde{G}}$ on $%
\mathrm{F}_{-}$.

\section{Complex product structure on Lie algebras}

A \emph{quadratic Lie algebra} is a Lie algebra $\mathfrak{g}$ equipped with
an invariant, nondegenerate symmetric bilinear form $\left( ,\right) _{%
\mathfrak{g}}$. Let $\mathfrak{g}$ be a quadratic Lie algebra with a split
bilinear form $\left( ,\right) _{\mathfrak{g}}$, and assume that the
underlying vector space is supplied with a double decomposition so it has
associated a pair of operators $\mathcal{E},\mathcal{J}:\mathfrak{g}%
\longrightarrow \mathfrak{g}$ such that $\mathcal{E}^{2}=\mathcal{I}$, $%
\mathcal{J}^{2}=-\mathcal{I}$ and $\mathcal{EJ}+\mathcal{JE}=0$. In presence
of a Lie algebra structure, it is important to pay attention to
integrability issues: on a Lie algebra $\mathfrak{g}$ a linear operator $%
\mathcal{E}:\mathfrak{g}\longrightarrow \mathfrak{g}$ satisfying $\mathcal{E}%
^{2}=\mathcal{I}$ is called an \emph{almost product} \emph{structure}, and
it is said integrable if the Nijehuis condition is satisfied, namely 
\begin{equation*}
\left[ \mathcal{E}X,\mathcal{E}Y\right] -\mathcal{E}\left( \left[ \mathcal{E}%
X,Y\right] +\left[ X,\mathcal{E}Y\right] \right) +\left[ X,Y\right] =0
\end{equation*}%
for all $X,Y\in \mathfrak{g}$. Equivalently, the linear operator $\mathcal{E}
$ is integrable iff its eigenspaces $\mathrm{E}_{+}$ and $\mathrm{E}_{-}$
are Lie subalgebras of $\mathfrak{g}$. An integrable almost product
structure is called a \emph{product} \emph{structure}. If the eigenspaces $%
\mathrm{E}_{+}$ and $\mathrm{E}_{-}$, associated with the eigenvalues $+1$
and $-1$, respectively, have the same dimension, the product structure is
called a \emph{paracomplex} \emph{structure}. \cite{Liberman}

A linear operator $\mathcal{J}:\mathfrak{g}\longrightarrow \mathfrak{g}$
satisfying $\mathcal{J}^{2}=-\mathcal{I}$ is called an \emph{almost complex
structure }and it is integrable if the Nijenhuis condition%
\begin{equation*}
\left[ \mathcal{J}X,\mathcal{J}Y\right] -\mathcal{J}\left( \left[ \mathcal{J}%
X,Y\right] +\left[ X,\mathcal{J}Y\right] \right) -\left[ X,Y\right] =0
\end{equation*}%
is satisfied. In this case, it is called a \emph{complex} \emph{structure}.

A \emph{complex product structure} on a Lie algebra $\mathfrak{g}$ is given
by a product structure $\mathcal{E}$ and a complex structure $\mathcal{J}$
such that $\mathcal{EJ}+\mathcal{JE}=0$. Complex product structures on Lie
algebras are exhaustively studied in ref. \cite{Andrada-Salamon}. In this
work we are involved with almost complex structures, and the product
structures become integrable only in some particular case.

Next we will apply the study of section \ref{cp vector spaces} to a Manin
triple endowed with a generalized metric. A Manin triple consists of a
triple of Lie algebras $\left( \mathfrak{g},\mathfrak{g}_{+},\mathfrak{g}%
_{-}\right) $ where $\mathfrak{g}$ is equipped with an invariant
nondegenerate symmetric bilinear form $\left( ,\right) _{\mathfrak{g}}$ and $%
\mathfrak{g}_{+},\mathfrak{g}_{-}$ are Lagrangian (maximally isotropic) Lie
subalgebras of $\mathfrak{g}$ such that $\mathfrak{g}=\mathfrak{g}_{+}\oplus 
\mathfrak{g}_{-}$. There is one-to-one correspondence between Manin triples
on $\mathfrak{g}$ and Lie bialgebra structures on the Lie algebra $\mathfrak{%
g}_{\pm }$ \cite{Lu-Weinstein}, which in turn are in one-to-one
correspondence with Poisson-Lie structures on the connected and
simply-connected Lie group $G_{\pm }$ associated with $\mathfrak{g}_{\pm }$.
The main idea behind the correspondence between Manin triples on $\mathfrak{g%
}$ and Lie bialgebra structures on the Lie algebra $\mathfrak{g}_{\pm }$ is
the identification $\mathfrak{g}_{\pm }\simeq \mathfrak{g}_{\mp }^{\ast }$
through the linear bijection induced by the bilinear form $\left( ,\right) _{%
\mathfrak{g}}$, such that $\mathfrak{g}=\mathfrak{g}_{+}\oplus \mathfrak{g}%
_{-}\simeq \mathfrak{g}_{\pm }\oplus \mathfrak{g}_{\pm }^{\ast }=\mathcal{D}%
\left( \mathfrak{g}_{\pm }\right) $, where $\mathcal{D}\left( \mathfrak{g}%
_{\pm }\right) $ is called the double Lie algebra of $\mathfrak{g}_{\pm }$,
or the classical double of $\mathfrak{g}_{\pm }$. In fact, a Lie bialgebra
on the Lie algebra $\mathfrak{g}_{\pm }$ is defined by a Lie cobracket $%
\delta :\mathfrak{g}_{\pm }\longrightarrow \mathfrak{g}_{\pm }\otimes 
\mathfrak{g}_{\pm }$ such that the dual map $\delta ^{\ast }:\mathfrak{g}%
_{\pm }^{\ast }\otimes \mathfrak{g}_{\pm }^{\ast }\longrightarrow \mathfrak{g%
}_{\pm }^{\ast }$ defines a Lie bracket on $\mathfrak{g}_{\pm }^{\ast }$.
There is also a compatibility condition between the Lie cobracket and the
Lie bracket on $\mathfrak{g}_{\pm }$ which implies that $\delta $ is $1$%
-cocycle with values in $\mathfrak{g}_{\pm }\otimes \mathfrak{g}_{\pm }$.
Then, building a cobracket from a coboundary $r\in \mathfrak{g}_{\pm
}\otimes \mathfrak{g}_{\pm }$ (the $r$-matrix) fulfills automatically this
requirement. It remains the co-Jacobi condition (or the Jacobi identity for
the induced Lie bracket in $\mathfrak{g}_{\pm }^{\ast }$) which is a hard
restriction on the $r$-matrix that boils down to asking for the
ad-invariance of the objects $r_{12}+r_{21}\in \mathfrak{g}_{\pm }\otimes 
\mathfrak{g}_{\pm }$ and $\left[ r_{13},r_{23}\right] +\left[ r_{12},r_{13}%
\right] +\left[ r_{12},r_{23}\right] \in \mathfrak{g}_{\pm }\otimes 
\mathfrak{g}_{\pm }\otimes \mathfrak{g}_{\pm }$, giving rise to the
classical Yang-Baxter equation and the modified classical Yang-Baxter
equation.

\section{From Manin triples to Lie algebras orthogonal direct sum}

Starting from a Manin triple $\left( \mathfrak{g},\mathfrak{g}_{+},\mathfrak{%
g}_{-}\right) $, that is a quadratic Lie algebra, and a generalized metric $%
\mathcal{G}+\mathcal{B}:\mathfrak{g}_{+}\longrightarrow \mathfrak{g}_{-}$ we
build a new Manin triple structure admitting an orthogonal direct sum
decomposition in terms of Lie algebra ideals associated with the graphs of $%
\mathcal{G}+\mathcal{B}$ and its orthogonal complement, namely the graph of $%
\mathcal{B}-\mathcal{G}$. This will be achieved by turning $\mathcal{B}$
into an $\mathcal{O}$-operator with extension $\mathcal{G}$ of mass\textit{\ 
}$\kappa =-1$.

\subsection{Generalized metrics and almost complex product structures}

Let $\left( \mathfrak{g},\mathfrak{g}_{+},\mathfrak{g}_{-}\right) $ be a
Manin triple with $\dim \mathfrak{g}=2n$. We assume that $\mathfrak{g}_{+}$
is equipped with a generalized metric realized through a linear map\textrm{\ 
}$\mathcal{H}:\mathfrak{g}_{+}\longrightarrow \mathfrak{g}_{-}$ such that,
relative to the bilinear form $\left( ,\right) _{\mathfrak{g}}$, it can be
decomposed as $\mathcal{H}=\mathcal{G}+\mathcal{B}$ where $\mathcal{G}$ is
the symmetric component, assumed to be invertible, and $\mathcal{B}$ is the
skew-symmetric one.

Let $\mathcal{H}^{\top }:\mathfrak{g}_{+}\longrightarrow \mathfrak{g}_{-}$
be the transpose of $\mathcal{H}:\mathfrak{g}_{+}\longrightarrow \mathfrak{g}%
_{-}$, $\left( \mathcal{H}X_{+},Y_{+}\right) _{\mathfrak{g}}=\left( X_{+},%
\mathcal{H}^{\top }Y_{+}\right) _{\mathfrak{g}}$, and consider the subspaces 
$\mathcal{E}^{+}$ and $\mathcal{E}^{-}$ of $\mathfrak{g}$ defined by the
graphs of the linear maps $\mathcal{H}$ and $-\mathcal{H}^{\top }$, 
\begin{equation}
\mathcal{E}^{\pm }=\left\{ X_{+}+\left( \mathcal{B}\pm \mathcal{G}\right)
X_{+}/X_{+}\in \mathfrak{g}_{+}\right\} .  \label{cpd 04c}
\end{equation}%
$\mathcal{E}^{+}$ and $\mathcal{E}^{-}$ are transversal orthogonal subspaces
of dimension $n$ such that $\mathfrak{g}=\mathcal{E}^{+}\oplus \mathcal{E}%
^{-}$, and the vector space $\mathfrak{g}$ admits a double splitting, namely 
$\mathfrak{g}=\mathfrak{g}_{+}\oplus \mathfrak{g}_{-}=\mathcal{E}^{+}\oplus 
\mathcal{E}^{-}$. We describe these orthogonal subspaces by introducing a
the linear operators $\mathcal{E},\mathcal{J}:\mathfrak{g}\longrightarrow 
\mathfrak{g}$ such that 
\begin{eqnarray*}
\mathcal{E}\left( X_{+}+\mathcal{B}X_{+}\pm \mathcal{G}X_{+}\right) &=&\pm
\left( X_{+}+\mathcal{B}X_{+}\pm \mathcal{G}X_{+}\right)  \notag \\
&& \\
\mathcal{J}\left( X_{+}+\mathcal{B}X_{+}\pm \mathcal{G}X_{+}\right) &=&\pm
\left( X_{+}+\mathcal{B}X_{+}\mp \mathcal{G}X_{+}\right)  \notag
\end{eqnarray*}%
implying that $\mathcal{E}$ and $\mathcal{J}$ are symmetric operators
satisfying the properties 
\begin{equation*}
\begin{array}{ccccc}
\mathcal{E}^{2}=\mathcal{I} & , & \mathcal{J}^{2}=-\mathcal{I} & , & 
\mathcal{EJ}+\mathcal{JE}=0%
\end{array}%
,
\end{equation*}%
that is, they conform an almost complex product structure on $\mathfrak{g}$. 
$\mathcal{J}$ is anti-compatible with the bilinear form $\left( ,\right) _{%
\mathfrak{g}}$, turning $\mathfrak{g}$ in a complex vector space with an
anti-Hermitian metric.

In the direct sum decomposition $\mathfrak{g}=\mathfrak{g}_{+}\oplus 
\mathfrak{g}_{-}$, the block matrix form of these operators are 
\begin{equation*}
\begin{array}{lll}
\mathcal{E}=\left( 
\begin{array}{cc}
-\mathcal{G}^{-1}\mathcal{B} & \mathcal{G}^{-1} \\ 
\mathcal{G}-\mathcal{BG}^{-1}\mathcal{B} & \mathcal{BG}^{-1}%
\end{array}%
\right) & , & \mathcal{J}=\left( 
\begin{array}{cc}
-\mathcal{G}^{-1}\mathcal{B} & \mathcal{G}^{-1} \\ 
-\mathcal{G}-\mathcal{BG}^{-1}\mathcal{B} & \mathcal{BG}^{-1}%
\end{array}%
\right)%
\end{array}%
\end{equation*}%
while in the orthogonal decomposition they take the form given in eq. $%
\left( \ref{cpd 05b}\right) $. They span a tridimensional algebra with basis 
$\left\{ \mathcal{E},\mathcal{J},\mathcal{F}\right\} $, where $\mathcal{F}=%
\mathcal{JE}$ is an involutive skew-symmetric operator. In the present
framework, the linear operators $\left\{ \mathcal{E},\mathcal{J}\right\} $
are not integrable in general.

\subsection{Generalized metrics and $\mathcal{O}$-operators}

The next step is to bring the generalized metric $\mathcal{H}=\mathcal{G}+%
\mathcal{B}$ on $\mathfrak{g}_{+}$ into the context of the $r$-matrix method 
\cite{Drinfeld-00},\cite{Semenov-01}. We consider the generalization of the $%
r$-matrix method introduced in \cite{Kosmann-Magri-01},\cite{Bordemann} and
the further extensions of ref. \cite{Chengming-01}-\cite{Chengming-02}, by
regarding the $r$-matrices as linear maps from a representation space of a
Lie algebra to the Lie algebra itself. In the last references, the method is
extended to the case in which the representation space is also a Lie algebra
and the action is a derivation. Although the setting of refs. \cite%
{Kosmann-Magri-01},\cite{Bordemann} would be enough for the current
developments, we will use the more general and versatile method of $\mathcal{%
O}$-operators of reference \cite{Chengming-01}, combined with the \emph{%
twilled} extension of Lie algebras of ref. \cite{Kosmann-Magri-01}.

Let's make a brief review of some basic definitions of the $\mathcal{O}$%
-operators limited to what is needed in this work. Let $\mathfrak{k}$ be a
finite dimensional Lie algebra and $\mathrm{V}$ a representation space of $%
\mathfrak{k}$, and denote by $\sigma :\mathfrak{k}\longrightarrow \mathrm{End%
}\left( \mathrm{V}\right) $ the linear map such that $X\longmapsto \sigma
_{X}$ for $X\in \mathfrak{k}$. A linear map $\beta :\mathrm{V}%
\longrightarrow \mathfrak{k}$ is said \emph{antisymmetric }if%
\begin{equation*}
\sigma _{\beta \left( V\right) }W+\sigma _{\beta \left( W\right) }V=0
\end{equation*}%
for $V,W\in \mathrm{V}$, and it is said $\mathfrak{k}$-\emph{invariant} if 
\begin{equation*}
\beta \left( \sigma _{X}V\right) =\left[ X,\beta \left( V\right) \right]
\end{equation*}%
for $X\in \mathfrak{k},Y\in \mathrm{V}$. Then, if $\beta :\mathrm{V}%
\longrightarrow \mathfrak{k}$ be a linear map, antisymmetric,\textit{\ }$%
\mathfrak{k}$\textit{-}invariant, a linear map $r:\mathrm{V}\longrightarrow 
\mathfrak{k}$ is called an\textit{\ }\emph{extended}\textit{\ }$\mathcal{O}$%
\textit{-}\emph{operator}\textit{\ }\emph{with}\textit{\ }\emph{extension} $%
\beta $ \emph{of}\textit{\ }\emph{mass}\textit{\ }$\left( -1\right) $ if it
fulfills the equation\textit{\ }%
\begin{equation}
\left[ r\left( V\right) ,r\left( W\right) \right] -r\left( \sigma _{r\left(
V\right) }W-\sigma _{r\left( W\right) }V\right) =-\left[ \beta \left(
V\right) ,\beta \left( W\right) \right]  \label{ybo 06f}
\end{equation}%
for $V,W\in \mathrm{V}$.

The main result for our purpose is established in the following theorem
(Theorem 2.18 in ref. \cite{Bordemann}, and a restricted version of Theorem
2.13 in ref. \cite{Chengming-01}).

\begin{description}
\item[Theorem] \textit{Let} $\mathfrak{k}$ \textit{be a finite dimensional
Lie algebra, }$\mathrm{V}$ \textit{be a} \textit{a representation space} 
\textit{of} $\mathfrak{k}$\textit{, }$r,\beta :\mathrm{V}\rightarrow 
\mathfrak{k}$ \textit{be linear maps.}

\begin{enumerate}
\item[i.] \textit{If} $r:\mathrm{V}\rightarrow \mathfrak{k}$ \textit{is an
extended }$\mathcal{O}$\textit{-operator with extension} $\beta $ \textit{of
mass }$\kappa $\textit{, then the bracket}%
\begin{equation*}
\left[ V,W\right] _{r}=\sigma _{r\left( V\right) }W-\sigma _{r\left(
W\right) }V
\end{equation*}%
\textit{defines a Lie algebra on} $\mathrm{V}$. \textit{We denote this Lie
algebra} $\left( \mathrm{V},\left[ ,\right] _{r}\right) $ \textit{as} $%
\mathrm{V}^{r}$.

\item[ii.] \textit{If} $\beta :\mathrm{V}\rightarrow \mathfrak{k}$ \textit{%
is }$\mathfrak{k}$-\textit{invariant of mass }$\kappa =$-\textit{1, then} $r$
\textit{satisfies }%
\begin{equation*}
\left[ r\left( V\right) ,r\left( W\right) \right] -r\left( \sigma _{r\left(
V\right) }W-\sigma _{r\left( W\right) }V\right) =-\left[ \beta \left(
V\right) ,\beta \left( W\right) \right]
\end{equation*}%
\textit{if and only if }$\left( r\pm \beta \right) :\mathrm{V}%
^{r}\longrightarrow \mathfrak{k}$ \textit{is a Lie algebra homomorphism,
namely} \textit{\ }%
\begin{equation*}
\left( r\pm \beta \right) \left[ V,W\right] _{r}=\left[ \left( r\pm \beta
\right) V,\left( r\pm \beta \right) W\right]
\end{equation*}%
$\forall V,W\in $ $\mathrm{V}^{r}$.
\end{enumerate}
\end{description}

Equation $\left( \ref{ybo 06f}\right) $ comes to play the role of the
modified classical Yang-Baxter equation. Under these conditions, the new Lie
bracket on $\mathrm{V}$ defines the coboundary cobracket on $\delta _{r}$ on 
$\mathfrak{k}$.

\subsection{\label{B as an O-operator}$\mathcal{B}$ as an extended $\mathcal{%
O}$-operator with extension $\mathcal{G}$}

Let's start with a Manin triple $\left( \mathfrak{g},\mathfrak{g}_{+},%
\mathfrak{g}_{-}\right) $ with $\dim \mathfrak{g}=$ $2n$ equipped with an
almost product structure $\mathcal{E}$ and an almost complex structure $%
\mathcal{J}$ such that $\mathcal{EJ}+\mathcal{JE}=0$ so, besides the
Lagrangian splitting $\mathfrak{g}=\mathfrak{g}_{+}\oplus \mathfrak{g}_{-}$,
it admits a vector space orthogonal splitting $\mathfrak{g}=\mathcal{E}%
^{+}\oplus \mathcal{E}^{-}$. Now, we will use the linear operators $\mathcal{%
G},\mathcal{B}:\mathfrak{g}_{+}\longrightarrow \mathfrak{g}_{-}$ to build a
new Lie bialgebra structure on $\mathfrak{g}_{+}$ by asking for $\mathcal{B}$
to be an extended $\mathcal{O}$-operator with extension $\mathcal{G}$ of mass%
\textit{\ }$\kappa =-1$. In doing so, we consider the vector space $%
\mathfrak{g}_{+}$ as a representation space of $\mathfrak{g}_{-}$ by the map 
$\sigma :\mathfrak{g}_{-}\longrightarrow \mathrm{End}\left( \mathfrak{g}%
_{+}\right) /X_{-}\longmapsto \sigma _{X_{-}}$ defined as 
\begin{equation}
\sigma _{X_{-}}Y_{+}=\Pi _{\mathfrak{g}_{+}}\left[ X_{-},Y_{+}\right]
\label{Bop 00}
\end{equation}%
where $\Pi _{\mathfrak{g}_{\pm }}:\mathfrak{g}\longrightarrow \mathfrak{g}%
_{\pm }$ is the projector. It is just the dressing action of $\mathfrak{g}%
_{-}$ on $\mathfrak{g}_{+}$. In addition we assume that $\mathcal{G}$ is $%
\mathfrak{g}_{-}$-invariant which means that%
\begin{equation}
\mathcal{G}\left( \sigma _{X_{-}}Y_{+}\right) =\left[ X_{-},\mathcal{G}Y_{+}%
\right]  \label{Bop 01}
\end{equation}%
for $X_{-}\in \mathfrak{g}_{-}$ and $Y_{+}\in \mathfrak{g}_{+}$. It also
implies that $\mathcal{G}$ is antisymmetric 
\begin{equation*}
\sigma _{\mathcal{G}X_{+}}Y_{+}+\sigma _{\mathcal{G}Y_{+}}X_{+}=0.
\end{equation*}%
Then we take $\mathcal{B}$ as an extended $\mathcal{O}$-operator with
extension $\mathcal{G}$ of mass $\kappa =-1$, therefore $\mathcal{B}$ and $%
\mathcal{G}$ fulfill the condition 
\begin{equation}
\left[ \mathcal{B}X_{+},\mathcal{B}Y_{+}\right] -\mathcal{B}\left( \sigma _{%
\mathcal{B}X_{+}}Y_{+}-\sigma _{\mathcal{B}Y_{+}}X_{+}\right) =-\left[ 
\mathcal{G}X_{+},\mathcal{G}Y_{+}\right]  \label{Bop 02a}
\end{equation}%
for $X_{+},Y_{+}\in \mathfrak{g}_{+}$. In turn, this implies that the bracket%
\begin{equation}
\left[ X_{+},Y_{+}\right] _{\mathcal{B}}=\sigma _{\mathcal{B}%
X_{+}}Y_{+}-\sigma _{\mathcal{B}Y_{+}}X_{+}  \label{Bop 03a}
\end{equation}%
defines a new Lie algebra structure $\left( \mathfrak{g}_{+},\left[ ,\right]
_{\mathcal{B}}\right) $ and eq. $\left( \ref{Bop 02a}\right) $ is equivalent
to stating that $\left( \mathcal{B}\pm \mathcal{G}\right) :\mathfrak{g}_{+}^{%
\mathcal{B}}\longrightarrow \mathfrak{g}_{-}$ is a Lie algebra homomorphism 
\begin{equation}
\left( \mathcal{B}\pm \mathcal{G}\right) \left[ X_{+},Y_{+}\right] _{%
\mathcal{B}}=\left[ \left( \mathcal{B}\pm \mathcal{G}\right) X_{+},\left( 
\mathcal{B}\pm \mathcal{G}\right) Y_{+}\right]  \label{Bop 03b}
\end{equation}%
for $X_{+},Y_{+}\in $ $\left( \mathfrak{g}_{+},\left[ ,\right] _{\mathcal{B}%
}\right) $. We denote the Lie algebra $\left( \mathfrak{g}_{+},\left[ ,%
\right] _{\mathcal{B}}\right) $ as $\mathfrak{g}_{+}^{\mathcal{B}}$.

\subsection{\label{twilled extensions}Twilled extensions}

We shall use a \emph{twilled extension} of Lie algebras, introduced in
reference \cite{Kosmann-Magri-01}, to obtain a new Lie algebra structure
from $\mathfrak{g}_{-}$ and $\mathfrak{g}_{+}^{\mathcal{B}}$, so let's have
a brief review of how it works. Consider a couple of Lie algebras $\mathfrak{%
g}_{-}$ and $\mathfrak{g}_{+}$, equipped with linear maps $\sigma :\mathfrak{%
g}_{-}\longrightarrow \mathrm{End}\left( \mathfrak{g}_{+}\right)
/X_{-}\longmapsto \sigma _{X_{-}}$ and $\rho :\mathfrak{g}%
_{+}\longrightarrow \mathrm{End}\left( \mathfrak{g}_{-}\right)
/X_{+}\longmapsto \rho _{X_{-}}$ turning $\mathfrak{g}_{+}$ into a
representation space of $\mathfrak{g}_{-}$ and $\mathfrak{g}_{-}$ into a
representation space of $\mathfrak{g}_{+}$, respectively. With this input,
the twilled extension $\mathfrak{g}$ of $\mathfrak{g}_{-}$ and $\mathfrak{g}%
_{+}$ is the vector space $\mathfrak{g}_{+}\oplus \mathfrak{g}_{-}$ endowed
with the skew-symmetric bracket%
\begin{equation*}
\left[ X_{+}+X_{-},Y_{+}+Y_{-}\right] _{\mathfrak{g}_{\mathcal{B}}}=\left[
X_{+},Y_{+}\right] +\sigma _{X_{-}}Y_{+}-\sigma _{Y_{-}}X_{+}+\left[
X_{-},Y_{-}\right] +\rho _{X_{+}}Y_{-}-\rho _{Y_{+}}X_{-}
\end{equation*}%
which is a Lie algebra structure on $\mathfrak{g}_{+}\oplus \mathfrak{g}_{-}$
iff $\rho $ and $\sigma $ satisfy a pair of constraints in order to ensure
the validity of the Jacobi identity. These constraints are most conveniently
expressed in terms of so called \emph{orbit maps }defined as the assignments%
\emph{\ }$Y_{-}\in \mathfrak{g}_{-}\longmapsto \rho _{Y_{-}}\in \mathrm{Hom}%
_{lin}\left( \mathfrak{g}_{+},\mathfrak{g}_{-}\right) $ and $Y_{+}\in 
\mathfrak{g}_{+}\longmapsto \sigma _{Y_{+}}\in \mathrm{Hom}_{lin}\left( 
\mathfrak{g}_{-},\mathfrak{g}_{+}\right) $ defined as 
\begin{equation*}
\begin{array}{ccc}
\rho _{Y_{-}}X_{+}:=\rho _{X_{+}}Y_{-} & \text{,} & \sigma
_{Y_{+}}X_{-}:=\sigma _{X_{-}}Y_{+}%
\end{array}%
.
\end{equation*}%
Thus the constraints take the form 
\begin{equation}
\begin{array}{l}
\rho _{\left[ X_{-},Y_{-}\right] }=ad_{X_{-}}\circ \rho _{Y_{-}}-\rho
_{Y_{-}}\circ \sigma _{X_{-}}+\rho _{X_{-}}\circ \sigma
_{Y_{-}}-ad_{Y_{-}}\circ \rho _{X_{-}} \\ 
\\ 
\sigma _{\left[ X_{+},Y_{+}\right] }=ad_{X_{+}}\circ \sigma _{Y_{+}}-\sigma
_{Y_{+}}\circ \rho _{X_{+}}+\sigma _{X_{+}}\circ \rho
_{Y_{+}}-ad_{Y_{+}}\circ \sigma _{X_{+}}%
\end{array}
\label{tw 04}
\end{equation}%
As shown in ref. \cite{Kosmann-Magri-01}, these relations mean that $\rho $
is $1$-cocycles on $\mathfrak{g}_{-}$ with values in $\mathrm{Hom}%
_{lin}\left( \mathfrak{g}_{+},\mathfrak{g}_{-}\right) $, and $\sigma $ is a $%
1$-cocycle on $\mathfrak{g}_{+}$ with values in $\mathrm{Hom}_{lin}\left( 
\mathfrak{g}_{-},\mathfrak{g}_{+}\right) $.

\subsubsection{The twilled extension of $\mathfrak{g}_{+}^{\mathcal{B}},%
\mathfrak{g}_{-}$}

Now we specialize this method to the setting of subsection \ref{B as an
O-operator} where $\mathfrak{g}_{+}$ is a representation space of $\mathfrak{%
g}_{-}$ by the map $\sigma :\mathfrak{g}_{-}\longrightarrow \mathrm{End}%
\left( \mathfrak{g}_{+}\right) /X_{-}\longmapsto \sigma _{X_{-}}$ defined in 
$\left( \ref{Bop 00}\right) $, and $\mathfrak{g}_{+}$ is endowed with the
Lie bracket $\left[ ,\right] _{\mathcal{B}}$ of eq. $\left( \ref{Bop 03a}%
\right) $. So, we need to define a map $\rho :\mathfrak{g}%
_{+}\longrightarrow \mathrm{End}\left( \mathfrak{g}_{-}\right)
/X_{+}\longmapsto \rho _{X_{-}}$ turning $\mathfrak{g}_{-}$ into a
representation space of $\mathfrak{g}_{+}$. Since $\sigma $ is given, we
focus on finding a linear map $\rho $ fulfilling the constraints $\left( \ref%
{tw 04}\right) $ following path of \emph{exact} twilled extension which
ensures the constraints are satisfied \cite{Kosmann-Magri-01}. Regarding $%
\mathrm{Hom}_{lin}\left( \mathfrak{g}_{+},\mathfrak{g}_{-}\right) $ as a $%
\mathfrak{g}_{-}$-module under the left action $\mathfrak{g}_{-}\times 
\mathrm{Hom}_{lin}\left( \mathfrak{g}_{+},\mathfrak{g}_{-}\right)
\longrightarrow \mathrm{Hom}_{lin}\left( \mathfrak{g}_{+},\mathfrak{g}%
_{-}\right) $ defined as 
\begin{equation*}
\left( X_{-},\rho _{Y_{-}}\right) \longmapsto X_{-}\cdot \rho _{Y_{-}}=\rho
_{Y_{-}}\circ \sigma _{X_{-}}-ad_{X_{-}}\circ \rho _{Y_{-}}
\end{equation*}%
for $\left( X_{-},\rho _{Y_{-}}\right) \in $ $\mathfrak{g}_{-}\times \mathrm{%
Hom}_{lin}\left( \mathfrak{g}_{+},\mathfrak{g}_{-}\right) $, a $1$-form $%
\rho $ on $\mathfrak{g}_{-}$ with values in the $\mathfrak{g}_{-}$-module $%
\mathrm{Hom}_{lin}\left( \mathfrak{g}_{+},\mathfrak{g}_{-}\right) $ is a $1$%
-cocycle if 
\begin{equation*}
d\rho \left( X_{-},Y_{-}\right) =X_{-}\cdot \rho _{Y_{-}}-\,Y_{-}\cdot \rho
_{X_{-}}-\rho _{\lbrack X_{-},Y_{-}]}=0
\end{equation*}%
which is equivalent to the first constraint in eq. $\left( \ref{tw 04}%
\right) $. An obvious solution to the first constraint is to choose a map $%
\rho $ as a coboundary, i.e., we take $\rho =d\mathcal{B}$ with $\mathcal{B}%
\in \mathrm{Hom}_{lin}\left( \mathfrak{g}_{+},\mathfrak{g}_{-}\right) $
regarded as a $0$-form, then 
\begin{equation}
\rho _{X_{-}}=d\mathcal{B}\left( X_{-}\right) =\mathcal{B}\circ \sigma
_{X_{-}}-ad_{X_{-}}\circ \mathcal{B}.  \label{tw 09}
\end{equation}%
and it produces the linear action 
\begin{equation*}
\rho _{Y_{+}}X_{-}=\mathcal{B}\sigma _{X_{-}}Y_{+}-\left[ X_{-},\mathcal{B}%
Y_{+}\right]
\end{equation*}%
Moreover, the second constraint in eq. $\left( \ref{tw 04}\right) $ reduce
to a trivial identity. Thus, we get a \emph{right exact twilled extension} $%
\mathfrak{g}_{\mathcal{B}}$ of $\mathfrak{g}_{+}^{\mathcal{B}}$ and $%
\mathfrak{g}_{-}$, namely the vector space $\mathfrak{g}_{+}\oplus \mathfrak{%
g}_{-}$ endowed with the Lie bracket%
\begin{eqnarray*}
\left[ X_{+}+X_{-},Y_{+}+Y_{-}\right] _{\mathfrak{g}_{\mathcal{B}}} &=&\left[
X_{+},Y_{+}\right] _{\mathcal{B}}+\sigma _{X_{-}}Y_{+}-\sigma _{Y_{-}}X_{+}+%
\left[ X_{-},Y_{-}\right] \\
&&+\mathcal{B}\sigma _{X_{-}}Y_{+}-\left[ X_{-},\mathcal{B}Y_{+}\right] -%
\mathcal{B}\sigma _{Y_{-}}X_{+}-\left[ Y_{-},\mathcal{B}X_{+}\right] .
\end{eqnarray*}

\subsubsection{The twilled extension of $\left( \mathfrak{g}_{+}^{\mathcal{B}%
}\right) ^{op},\mathfrak{g}_{-}$ and the orthogonal subspace $\mathcal{E}%
^{\pm }$ as Lie ideals}

There is a second twilled construction in the framework of the previous
subsection steaming from the following observation: if $\left( \sigma ,\rho
\right) $ fulfill the conditions $\left( \ref{tw 04}\right) $ for the Lie
algebras $\mathfrak{g}_{-}$ and $\mathfrak{g}_{+}^{\mathcal{B}}$, then $%
\left( \sigma ,-\rho \right) $ fulfills the Jacobi conditions\textit{\ }for
the Lie algebras $\mathfrak{g}_{-}$ and $\left( \mathfrak{g}_{+}\right)
^{op} $. Thus we get a right exact twilled extension $\mathfrak{\tilde{g}}_{%
\mathcal{B}}$ of $\left( \mathfrak{g}_{+}^{\mathcal{B}}\right) ^{op}$ and $%
\mathfrak{g}_{-}$, namely $\mathfrak{\tilde{g}}_{\mathcal{B}}=\left( 
\mathfrak{g}_{+}^{\mathcal{B}}\right) ^{op}\oplus \mathfrak{g}_{-}$ with Lie
bracket%
\begin{eqnarray*}
\left[ X_{+}+X_{-},Y_{+}+Y_{-}\right] _{\mathfrak{\tilde{g}}_{\mathcal{B}}}
&=&-\left[ X_{+},Y_{+}\right] _{\mathcal{B}}+\sigma _{X_{-}}Y_{+}-\sigma
_{Y_{-}}X_{+}+\left[ X_{-},Y_{-}\right] \\
&&+\mathcal{B}\sigma _{X_{-}}Y_{+}-\left[ X_{-},\mathcal{B}Y_{+}\right] -%
\mathcal{B}\sigma _{Y_{-}}X_{+}+\left[ Y_{-},\mathcal{B}X_{+}\right] .
\end{eqnarray*}%
Recalling that $\sigma :\mathfrak{g}_{-}\longrightarrow \mathrm{End}\left( 
\mathfrak{g}_{+}\right) $ was given in eq. $\left( \ref{Bop 00}\right) $,
the map $\rho :\left( \mathfrak{g}_{+}^{\mathcal{B}}\right) ^{op}\otimes 
\mathfrak{g}_{-}\longrightarrow \mathfrak{g}_{-}$ turns in 
\begin{equation*}
\rho \left( X_{+},Y_{-}\right) =\mathcal{B}\Pi _{\mathfrak{g}_{+}}\left[
X_{+},Y_{-}\right] -\left[ \mathcal{B}X_{+},Y_{-}\right]
\end{equation*}%
and the Lie bracket in $\mathfrak{\tilde{g}}_{\mathcal{B}}=\left( \mathfrak{g%
}_{+}^{\mathcal{B}}\right) ^{op}\oplus \mathfrak{g}_{-}$ gets the explicit
form 
\begin{eqnarray}
&&\left[ X_{+}+X_{-},Y_{+}+Y_{-}\right] _{\mathfrak{\tilde{g}}_{\mathcal{B}}}
\label{Bop 06b} \\
&=&-\left[ X_{+},Y_{+}\right] _{\mathcal{B}}+\Pi _{\mathfrak{g}_{+}}\left[
X_{-},Y_{+}\right] -\Pi _{\mathfrak{g}_{+}}\left[ Y_{-},X_{+}\right]  \notag
\\
&&+\left[ X_{-},Y_{-}\right] +\mathcal{B}\Pi _{\mathfrak{g}_{+}}\left[
X_{+},Y_{-}\right] -\left[ \mathcal{B}X_{+},Y_{-}\right] -\mathcal{B}\Pi _{%
\mathfrak{g}_{+}}\left[ Y_{+},X_{-}\right] +\left[ \mathcal{B}Y_{+},X_{-}%
\right] .  \notag
\end{eqnarray}%
This double Lie algebra has a remarkable property, which is expressed in the
following theorem.

\begin{description}
\item[Theorem:] \textit{The orthogonal subspaces }$\mathcal{E}^{+}$ \textit{%
and} $\mathcal{E}^{-}defined$\textit{\ in eq.} $\left( \ref{cpd 04c}\right) $
\textit{are Lie ideals in} $\mathfrak{\tilde{g}}_{\mathcal{B}}$\textit{, and
they are (anti) isomorphic to }$\mathfrak{g}_{-}$.
\end{description}

\textbf{Proof: }To prove that $\mathcal{E}^{\pm }$ is a Lie subalgebra of $%
\mathfrak{\tilde{g}}_{\mathcal{B}}$, we note that the $\mathfrak{g}_{-}$%
-component of the Lie bracket on $\mathcal{E}^{\pm }$ can be written as%
\begin{eqnarray*}
&&\Pi _{\mathfrak{g}_{-}}\left[ X_{+}+\left( \mathcal{B}\pm \mathcal{G}%
\right) X_{+},Y_{+}+\left( \mathcal{B}\pm \mathcal{G}\right) Y_{+}\right] _{%
\mathfrak{\tilde{g}}_{\mathcal{B}}} \\
&=&-\left[ \left( \mathcal{B}\pm \mathcal{G}\right) X_{+},\left( \mathcal{B}%
\pm \mathcal{G}\right) Y_{+}\right] +\mathcal{B}\sigma _{\left( \mathcal{B}%
\pm \mathcal{G}\right) X_{+}}Y_{+}-\mathcal{B}\sigma _{\left( \mathcal{B}\pm 
\mathcal{G}\right) Y_{+}}X_{+} \\
&&\pm \left[ \left( \mathcal{B}\pm \mathcal{G}\right) X_{+},\mathcal{G}Y_{+}%
\right] \mp \left[ \left( \mathcal{B}\pm \mathcal{G}\right) Y_{+},\mathcal{G}%
X_{+}\right]
\end{eqnarray*}%
and, by virtue of the $\mathfrak{g}_{-}$-symmetry $\left( \ref{Bop 01}%
\right) $ applied in the last line and because of the relation $\left( \ref%
{Bop 03b}\right) $, it reduces to 
\begin{eqnarray*}
&&\Pi _{\mathfrak{g}_{-}}\left[ X_{+}+\left( \mathcal{B}\pm \mathcal{G}%
\right) X_{+},Y_{+}+\left( \mathcal{B}\pm \mathcal{G}\right) Y_{+}\right] _{%
\mathfrak{\tilde{g}}_{\mathcal{B}}} \\
&=&\left( \mathcal{B}\pm \mathcal{G}\right) \left( \Pi _{\mathfrak{g}_{+}}%
\left[ X_{+}+\left( \mathcal{B}\pm \mathcal{G}\right) X_{+},Y_{+}+\left( 
\mathcal{B}\pm \mathcal{G}\right) Y_{+}\right] _{\mathfrak{\tilde{g}}_{%
\mathcal{B}}}\right)
\end{eqnarray*}%
Therefore, the full Lie bracket between elements of $\mathcal{E}^{\pm }$ can
be written as%
\begin{eqnarray*}
&&\left[ \left( \emph{I}+\mathcal{B}\pm \mathcal{G}\right) X_{+},\left( 
\emph{I}+\mathcal{B}\pm \mathcal{G}\right) Y_{+}\right] _{\mathfrak{\tilde{g}%
}_{\mathcal{B}}} \\
&=&\left( \emph{I}+\mathcal{B}\pm \mathcal{G}\right) \left( \Pi _{\mathfrak{g%
}_{+}}\left[ \left( \emph{I}+\mathcal{B}\pm \mathcal{G}\right) X_{+},\left( 
\emph{I}+\mathcal{B}\pm \mathcal{G}\right) Y_{+}\right] _{\mathfrak{\tilde{g}%
}_{\mathcal{B}}}\right)
\end{eqnarray*}%
proving that $\mathcal{E}^{\pm }$ is a Lie subalgebra. After the evaluation
of the $\mathfrak{g}_{+}$-component of the Lie bracket we get%
\begin{equation}
\left[ \left( \emph{I}+\mathcal{B}\pm \mathcal{G}\right) X_{+},\left( \emph{I%
}+\mathcal{B}\pm \mathcal{G}\right) Y_{+}\right] _{\mathfrak{\tilde{g}}_{%
\mathcal{B}}}=\pm 2\left( \emph{I}+\mathcal{B}\pm \mathcal{G}\right) 
\mathcal{G}^{-1}\left[ \mathcal{G}X_{+},\mathcal{G}Y_{+}\right] .
\label{tw 11}
\end{equation}

To prove that $\mathcal{E}^{\pm }$ is an ideal we evaluate the crossed Lie
bracket between elements of $\mathcal{E}^{+}$and $\mathcal{E}^{-}$, where we
use again the $\mathfrak{g}_{-}$-invariance of $\mathcal{G}$ so that after
some handling we arrive to%
\begin{equation*}
\left[ X_{+}+\left( \mathcal{B}+\mathcal{G}\right) X_{+},Y_{+}+\left( 
\mathcal{B}-\mathcal{G}\right) Y_{+}\right] _{\mathfrak{\tilde{g}}_{\mathcal{%
B}}}=-\left[ \mathcal{B}X_{+},\mathcal{B}Y_{+}\right] -\left[ \mathcal{G}%
X_{+},\mathcal{G}Y_{+}\right] +\mathcal{B}\left[ X_{+},Y_{+}\right] _{%
\mathcal{B}}
\end{equation*}%
which vanishes because of the $\mathcal{O}$-operator relation $\left( \ref%
{Bop 02a}\right) $, showing that $\mathcal{E}^{\pm }$ is a Lie algebra ideal.

Note that each $X_{+}\in \mathfrak{g}_{+}$ can be written as $X_{+}=\mathcal{%
G}^{-1}X_{-}$ for some $X_{-}\in \mathfrak{g}_{-}$, then eq. $\left( \ref{tw
11}\right) $ can be written as\textbf{\ }%
\begin{equation*}
\left[ \frac{1}{2}\left( I+\mathcal{B}\pm \mathcal{G}\right) \mathcal{G}%
^{-1}X_{-},\frac{1}{2}\left( \mathcal{I}+\mathcal{B}\pm \mathcal{G}\right) 
\mathcal{G}^{-1}Y_{-}\right] _{\mathfrak{\tilde{g}}_{\mathcal{B}}}=\pm \frac{%
1}{2}\left( I+\mathcal{B}\pm \mathcal{G}\right) \mathcal{G}^{-1}\left[
X_{-},Y_{-}\right]
\end{equation*}%
therefore $\left( \mathcal{I}+\mathcal{B}\pm \mathcal{G}\right) \mathcal{G}%
^{-1}/2:\mathfrak{g}_{-}\longrightarrow \mathcal{E}^{\pm }$ is a Lie algebra
(anti) homomorphism with a trivial kernel and, because $\mathfrak{g}_{-}$
and $\mathcal{E}^{\pm }$ have the same dimension, it is a bijection.$%
\blacksquare $

\begin{description}
\item[Corollary] \textit{The Lie algebra} $\mathfrak{\tilde{g}}_{\mathcal{B}%
} $ \textit{admits the double splitting} 
\begin{equation*}
\mathfrak{\tilde{g}}_{\mathcal{B}}=\left( \mathfrak{g}_{+}\right)
^{op}\oplus \mathfrak{g}_{-}=\mathcal{E}^{+}\oplus \mathcal{E}^{-}.
\end{equation*}
\end{description}

This Lie bracket leaves invariant the bilinear form $\left( ,\right) _{%
\mathfrak{g}}$ 
\begin{equation*}
\left( \left[ X_{+}+X_{-},Y_{+}+Y_{-}\right] _{\mathfrak{\tilde{g}}_{%
\mathcal{B}}},Z_{+}+Z_{-}\right) _{\mathfrak{\tilde{g}}}=-\left( Y_{+}+Y_{-},%
\left[ X_{+}+X_{-},Y_{+}+Y_{-}\right] _{\mathfrak{\tilde{g}}_{\mathcal{B}%
}}\right) _{\mathfrak{g}}
\end{equation*}%
so we get the following result.

\begin{description}
\item[Proposition] $\left( \mathfrak{\tilde{g}}_{\mathcal{B}},\left( 
\mathfrak{g}_{+}^{\mathcal{B}}\right) ^{op},\mathfrak{g}_{-}\right) $ 
\textit{equipped with} \textit{the bilinear form }$\left( ,\right) _{%
\mathfrak{g}}$ \textit{is a Manin triple.}
\end{description}

It is worth highlighting that the above results imply that the product
structure $\mathcal{E}$ is integrable, so it is a para-complex structure on $%
\mathfrak{\tilde{g}}_{\mathcal{B}}$. On the other hand, the non-abelian
character of the ideals $\mathcal{E}^{+}$ and $\mathcal{E}^{-}$ prevents the
integrability of the linear operator $\mathcal{J}$, other wise $\mathcal{E}%
^{+}$ and $\mathcal{E}^{-}$ would be abelian ideals (see ref. \cite%
{Andrada-Salamon}).

In reference to the operator $\mathcal{J}$, we can recover the
antiisomorphism $\varphi _{\mathcal{B}}:\mathcal{E}^{+}\longrightarrow 
\mathcal{E}^{-}$, see eq. $\left( \ref{cpd 07b}\right) $, as the composition 
\begin{equation*}
\varphi _{\mathcal{B}}=\frac{1}{2}\left( I+\mathcal{B}\mp \mathcal{G}\right) 
\mathcal{G}^{-1}\circ \left( \frac{1}{2}\left( I+\mathcal{B}\pm \mathcal{G}%
\right) \mathcal{G}^{-1}\right) ^{-1},
\end{equation*}%
with 
\begin{equation*}
\varphi _{\mathcal{B}}\left( X_{+}+\left( \mathcal{B}+\mathcal{G}\right)
X_{+}\right) =X_{+}+\left( \mathcal{B}-\mathcal{G}\right) X_{+}.
\end{equation*}%
It is interesting to note the relation between the Lie bracket $\left( \ref%
{tw 11}\right) $ and the integrability of the complex structure $\mathcal{J_{%
\mathcal{B}}}$. In fact, since in the $\mathcal{E}^{+}\oplus \mathcal{E}^{-}$
splitting, $\mathcal{J_{\mathcal{B}}}$ takes the form%
\begin{equation*}
\mathcal{J_{\mathcal{B}}}=\left( 
\begin{array}{cc}
0 & -\varphi _{\mathcal{B}}^{-1} \\ 
\varphi _{\mathcal{B}} & 0%
\end{array}%
\right)
\end{equation*}%
then, the Nijenhuis tensor reduce to \cite{Andrada-Salamon}, 
\begin{eqnarray*}
\mathcal{N}_{\varphi }\left( X^{+},Y^{+}\right) &=&\varphi _{\mathcal{B}}%
\left[ X^{+},Y^{+}\right] _{\mathfrak{\tilde{g}}_{\mathcal{B}}}+\varphi _{%
\mathcal{B}}^{-1}\left[ \varphi _{\mathcal{B}}\left( X^{+}\right) ,\varphi _{%
\mathcal{B}}\left( Y^{+}\right) \right] _{\mathfrak{\tilde{g}}_{\mathcal{B}}}
\\
&&-\left[ \varphi _{\mathcal{B}}\left( X^{+}\right) ,Y^{+}\right] _{%
\mathfrak{\tilde{g}}_{\mathcal{B}}}-\left[ X^{+},\varphi _{\mathcal{B}%
}\left( Y^{+}\right) \right] _{\mathfrak{\tilde{g}}_{\mathcal{B}}},
\end{eqnarray*}%
for $X^{+}=\left( \emph{I}+\mathcal{B}\pm \mathcal{G}\right)
X_{+},Y^{+}=\left( \emph{I}+\mathcal{B}\pm \mathcal{G}\right) Y_{+}\in 
\mathcal{E}^{+}$. Then, after some computations, we get 
\begin{equation*}
\mathcal{N}_{\varphi }\left( \left( \emph{I}+\mathcal{B}\pm \mathcal{G}%
\right) X_{+},\left( \emph{I}+\mathcal{B}\pm \mathcal{G}\right) Y_{+}\right)
=-4\left[ \mathcal{G}X_{+},\mathcal{G}Y_{+}\right]
\end{equation*}%
and the Lie bracket $\left( \ref{tw 11}\right) $ can be written as 
\begin{equation*}
\left[ X^{+},Y^{+}\right] _{\mathfrak{\tilde{g}}_{\mathcal{B}}}=\mp \frac{1}{%
2}\left( \mathcal{I}+\mathcal{B}\pm \mathcal{G}\right) \mathcal{G}^{-1}%
\mathcal{N}_{\varphi }\left( X^{+},Y^{+}\right)
\end{equation*}%
making clear that $\mathcal{J}$ is integrable iff $\mathcal{E}^{\pm }$ is
abelian.

\begin{description}
\item[Lemma] \textit{The linear map} $\varphi \mathcal{_{\mathcal{B}}}:%
\mathcal{E}^{+}\longrightarrow \mathcal{E}^{-}$ \textit{defining the almost
complex structure} $\mathcal{J}$ \textit{is a Lie algebra antihomomorphism.}
\end{description}

\textbf{Proof: }Let $X^{+}=X_{+}+\left( \mathcal{B}+\mathcal{G}\right) X_{+}$
and $Y^{+}=Y_{+}+\left( \mathcal{B}+\mathcal{G}\right) Y_{+}\ $in $\mathcal{E%
}^{+}$, then 
\begin{eqnarray*}
\varphi \mathcal{_{\mathcal{B}}}\left( \left[ X^{+},Y^{+}\right] _{\mathfrak{%
\tilde{g}}_{\mathcal{B}}}\right) &=&2\left( \mathcal{G}^{-1}\left[ \mathcal{G%
}X_{+},\mathcal{G}Y_{+}\right] +\left( \mathcal{B}-\mathcal{G}\right) 
\mathcal{G}^{-1}\left[ \mathcal{G}X_{+},\mathcal{G}Y_{+}\right] \right) \\
&=&-\left[ X_{+}+\left( \mathcal{B}-\mathcal{G}\right) X_{+},Y_{+}+\left( 
\mathcal{B}-\mathcal{G}\right) Y_{+}\right] _{\mathfrak{\tilde{g}}_{\mathcal{%
B}}}
\end{eqnarray*}%
then 
\begin{equation*}
\varphi \mathcal{_{\mathcal{B}}}\left( \left[ X^{+},Y^{+}\right] _{\mathfrak{%
\tilde{g}}_{\mathcal{B}}}\right) =-\left[ \varphi \mathcal{_{\mathcal{B}}}%
\left( X^{+}\right) ,\varphi \mathcal{_{\mathcal{B}}}\left( Y^{+}\right) %
\right] _{\mathfrak{\tilde{g}}_{\mathcal{B}}}.
\end{equation*}%
as stated above.$\blacksquare $

This property can be also expressed as 
\begin{equation*}
\mathcal{E}\left[ \mathcal{J}X,\mathcal{J}Y\right] _{\mathfrak{\tilde{g}}_{%
\mathcal{B}}}=\mathcal{J}\left[ X,Y\right] _{\mathfrak{\tilde{g}}_{\mathcal{B%
}}}.
\end{equation*}

\subsection{$\mathcal{E}^{\pm }$ as $\mathfrak{g}_{-}$-Lie algebras}

It is interesting to note that the Lie algebra $\mathfrak{g}_{-}$ is a
common component in the Manin triples $\left( \mathfrak{g},\mathfrak{g}_{+},%
\mathfrak{g}_{-}\right) $ and $\left( \mathfrak{\tilde{g}}_{\mathcal{B}},%
\mathfrak{g}_{+}^{\mathcal{B}},\mathfrak{g}_{-}\right) $, and acts as a
derivation on the subalgebras $\mathcal{E}^{+}$ and $\mathcal{E}^{-}$ of $%
\mathfrak{\tilde{g}}_{\mathcal{B}}$.

\begin{description}
\item[Proposition:] \textit{The Lie subalgebra} $\mathcal{E}^{\pm }\subset 
\mathfrak{\tilde{g}}_{\mathcal{B}}$ \textit{is a} $\mathfrak{g}_{-}$-\textit{%
Lie algebra under} \textit{the adjoint action in} $\mathfrak{g}_{\mathcal{B}%
} $.
\end{description}

\textbf{Proof: }Consider the Lie bracket $\left( \ref{Bop 06b}\right) $ of $%
\mathfrak{\tilde{g}}_{\mathcal{B}}$ restricted to $\mathfrak{g}_{-}\times 
\mathcal{E}^{\pm }$, then the adjoint action of $\mathfrak{g}_{-}$ on $%
\mathcal{E}^{\pm }$ is 
\begin{eqnarray*}
&&\mathrm{ad}_{X_{-}}^{\mathfrak{\tilde{g}}_{\mathcal{B}}}\left(
Y_{+}+\left( \mathcal{B}\pm \mathcal{G}\right) Y_{+}\right) \\
&=&\left[ X_{-},Y_{+}+\left( \mathcal{B}\pm \mathcal{G}\right) Y_{+}\right]
_{\mathfrak{\tilde{g}}_{\mathcal{B}}} \\
&=&\Pi _{\mathfrak{g}_{+}}\left[ X_{-},Y_{+}\right] +\left[ X_{-},\left( 
\mathcal{B}\pm \mathcal{G}\right) Y_{+}\right] \\
&&+\mathcal{B}\Pi _{\mathfrak{g}_{+}}\left[ X_{-},Y_{+}\right] +\left[ 
\mathcal{B}Y_{+},X_{-}\right]
\end{eqnarray*}%
and recalling that $\mathcal{G}:\mathfrak{g}_{+}\longrightarrow \mathfrak{g}%
_{-}$ is $\mathfrak{g}_{-}$-invariant, it reduces to

\begin{equation*}
\mathrm{ad}_{X_{-}}^{\mathfrak{\tilde{g}}_{\mathcal{B}}}\left( I+\mathcal{B}%
\pm \mathcal{G}\right) Y_{+}=\left( I+\mathcal{B}\pm \mathcal{G}\right) \Pi
_{\mathfrak{g}_{+}}\left[ X_{-},Y_{+}\right]
\end{equation*}%
therefore $\mathcal{E}^{\pm }$ are invariant subspaces under the adjoint
action of $\mathfrak{g}_{-}$\textit{. }Writing it in terms of the dressing
action of $\mathfrak{g}_{-}$ on $\mathfrak{g}_{+}$ in the Manin triple $%
\left( \mathfrak{g},\mathfrak{g}_{+},\mathfrak{g}_{-}\right) $, it turns in

\begin{equation*}
\mathrm{ad}_{X_{-}}^{\mathfrak{\tilde{g}}_{\mathcal{B}}}\left( I+\mathcal{B}%
\pm \mathcal{G}\right) Y_{+}=\left( I+\mathcal{B}\pm \mathcal{G}\right)
\left( Y_{+}\right) ^{X_{-}}
\end{equation*}%
which means that $\left( I+\mathcal{B}\pm \mathcal{G}\right) $ intertwines
between the dressing action of $\mathfrak{g}_{-}$ on $\mathfrak{g}_{+}$ and
the adjoint action of $\mathfrak{g}_{-}$ on $\mathcal{E}^{\pm }$.

Moreover, because of the Jacobi identity we get 
\begin{equation*}
\mathrm{ad}_{X_{-}}^{\mathfrak{\tilde{g}}_{\mathcal{B}}}\left[ X^{\pm
},Y^{\pm }\right] _{\mathfrak{\tilde{g}}_{\mathcal{B}}}=\left[ \mathrm{ad}%
_{X_{-}}^{\mathfrak{\tilde{g}}_{\mathcal{B}}}X^{\pm },Y^{\pm }\right] _{%
\mathfrak{\tilde{g}}_{\mathcal{B}}}+\left[ X^{\pm },\mathrm{ad}_{X_{-}}^{%
\mathfrak{\tilde{g}}_{\mathcal{B}}}Y^{\pm }\right] _{\mathfrak{\tilde{g}}_{%
\mathcal{B}}}
\end{equation*}%
therefore $\mathcal{E}^{\pm }$ are $\mathfrak{g}_{-}$-Lie algebras.$%
\blacksquare $

This action $\mathrm{ad}^{\mathfrak{\tilde{g}}_{\mathcal{B}}}:\mathfrak{g}%
_{-}\times \mathcal{E}^{\pm }\longmapsto \mathcal{E}^{\pm }$ can be promoted
to an action of the connected simply connected Lie group $G_{-}\subset 
\tilde{G}_{\mathcal{B}}$ associated with the Lie algebra $\mathfrak{g}_{-}$ $%
\mathrm{Ad}^{\tilde{G}_{\mathcal{B}}}:G_{-}\times \mathcal{E}^{\pm
}\longrightarrow \mathcal{E}^{\pm }$ /$\left( g_{-},Y^{\pm }\right)
\longmapsto \mathrm{Ad}_{g_{-}}^{\tilde{G}_{\mathcal{B}}}Y^{\pm }$ such that%
\begin{equation*}
\mathrm{Ad}_{g_{-}}^{\tilde{G}_{\mathcal{B}}}\left( I+\mathcal{B}\pm 
\mathcal{G}\right) Y_{+}=\left( I+\mathcal{B}\pm \mathcal{G}\right) \Pi _{%
\mathfrak{g}_{+}}\mathrm{Ad}_{g_{-}}^{G}Y_{+}
\end{equation*}%
meaning that the orbit of $G_{-}\subset \tilde{G}_{\mathcal{B}}$ through $%
X^{\pm }\in \mathcal{E}^{\pm }$ is the image of the dressing orbit of $%
G_{-}\subset G$ through $X_{+}\in \mathfrak{g}_{+}$ by the isomorphism $%
\left( I+\mathcal{B}\pm \mathcal{G}\right) :\mathfrak{g}_{+}\longrightarrow 
\mathcal{E}^{\pm }$.

\subsection{$\mathcal{B}$ and $\mathcal{G}$ from factorizable
quasitriangular $r$-matrices}

Let $\left( \mathfrak{g},\mathfrak{g}_{+},\mathfrak{g}_{-}\right) $ be a
Manin triple whose bilinear form is $\left( ,\right) _{\mathfrak{g}}$, and
let $r\in \mathfrak{g}_{-}\otimes \mathfrak{g}_{-}$ be a \emph{factorizable} 
\emph{quasitriangular }solution of the modified classical Yang-Baxter
equation, with $r^{+}$and $r^{-}$ the symmetric and skew-symmetric
components of $r$. Define the linear operators $\hat{r}^{-},\hat{r}^{+}:%
\mathfrak{g}_{-}^{\ast }\longrightarrow \mathfrak{g}_{-}$ as $\hat{r}^{\pm
}\left( \xi \right) =\left\langle \xi \otimes I,r^{\pm }\right\rangle $,
where $\hat{r}^{+}$ is invertible, then the $\mathfrak{g}_{-}$-invariance of 
$r^{+}$ becomes in 
\begin{equation*}
\hat{r}^{+}\circ ad_{X_{-}}^{\ast }+ad_{X_{-}}\circ \hat{r}^{+}=0.
\end{equation*}%
Here $ad_{X_{-}}$ denotes the adjoint action of $\mathfrak{g}_{-}$ on
itself, and in the whole work, $ad^{\ast }$ just means the transpose
operator, so the coadjoint action is the $-ad^{\ast }$.

By using the linear bijection $\psi :\mathfrak{g}_{+}\longrightarrow 
\mathfrak{g}_{-}^{\ast }$ induced by the bilinear form $\left( ,\right) _{%
\mathfrak{g}}$, we introduce the linear operators 
\begin{equation*}
\begin{array}{lll}
\mathcal{B}=\hat{r}^{-}\circ \psi :\mathfrak{g}_{+}\longrightarrow \mathfrak{%
g}_{-} & , & \mathcal{G}=\hat{r}^{+}\circ \psi :\mathfrak{g}%
_{+}\longrightarrow \mathfrak{g}_{-}%
\end{array}%
\end{equation*}%
then the $\mathfrak{g}_{-}$-invariance turns in 
\begin{equation}
\mathcal{G}\left( \Pi _{\mathfrak{g}_{+}}\left[ X_{-},Y_{+}\right] \right) =%
\left[ X_{-},\mathcal{G}Y_{+}\right] .  \label{ybo 08f}
\end{equation}%
The ad-invariance of the bilinear form means that $\psi ^{-1}\circ \mathrm{ad%
}_{X_{+}}^{\mathfrak{g}\ast }+\mathrm{ad}_{X_{+}}^{\mathfrak{g}}\circ \psi
^{-1}=0,$ therefore the action $\sigma :\mathfrak{g}_{-}\longrightarrow 
\mathrm{End}\left( \mathfrak{g}_{+}\right) $ introduced in eq. $\left( \ref%
{Bop 00}\right) $ can be written as 
\begin{equation*}
\sigma _{X_{-}}Y_{+}=\Pi _{\mathfrak{g}_{+}}\left[ X_{-},Y_{+}\right] =-\psi
^{-1}\left( ad_{X_{-}}^{\ast }\psi \left( Y_{+}\right) \right)
\end{equation*}%
then%
\begin{equation*}
\sigma _{\mathcal{B}X_{+}}Y_{+}=-\psi ^{-1}\left( ad_{\mathcal{B}%
X_{+}}^{\ast }\psi \left( Y_{+}\right) \right)
\end{equation*}%
or, equivalently, 
\begin{equation*}
\hat{r}^{-}ad_{\mathcal{B}X_{+}}^{\ast }\psi \left( Y_{+}\right) =-\mathcal{B%
}\Pi _{\mathfrak{g}_{+}}\left[ \mathcal{B}X_{+},Y_{+}\right] .
\end{equation*}%
This allows to write the classical Yang-Baxter equation%
\begin{equation*}
\left[ r_{13},r_{23}\right] +\left[ r_{12},r_{13}\right] +\left[
r_{12},r_{23}\right] =0
\end{equation*}%
as \cite{Bordemann} 
\begin{eqnarray*}
&&\left[ \mathcal{B}X_{+},\mathcal{B}Y_{+}\right] -\mathcal{B}\left( \sigma
\left( \mathcal{B}X_{+},Y_{+}\right) -\sigma \left( \mathcal{B}%
Y_{+},X_{+}\right) \right) \\
&&=-\left[ \mathcal{G}X_{+},\mathcal{G}Y_{+}\right]
\end{eqnarray*}%
so $\mathcal{B}$ is an $\mathcal{O}$-operator with extension $\mathcal{G}$
of mass\textit{\ }$-1$.

With the help of $\left( \ref{ybo 08f}\right) $ and writing $X_{+}=\mathcal{G%
}^{-1}X_{-}$ and $Y_{+}=\mathcal{G}^{-1}Y_{-}$, it turns in 
\begin{equation*}
[\mathcal{BG}^{-1}X_{-},\mathcal{BG}^{-1}Y_{-}]-\mathcal{BG}^{-1}\left[ 
\mathcal{BG}^{-1}X_{-},Y_{-}\right] -\mathcal{BG}^{-1}\left[ X_{-},\mathcal{%
BG}^{-1}Y_{-}\right] =-[X_{-},Y_{-}]
\end{equation*}%
showing that $\mathcal{BG}^{-1}$ is a solution of the modified classical
Yang-Baxter equation in the form of ref. \cite{Semenov-01}.

\bigskip

\subsection{Orthogonal splitting of a factorizable quasitriangular Lie
bialgebra}

Here we set aside the original Manin triple and apply the latter results on
a factorizable quasitriangular Lie bialgebra $\mathfrak{k}$, providing a
rather simple nontrivial example of the construction of this section. Thus,
the Lie bialgebra is defined by a quasitriangular factorizable $r$-matrix
which also plays the role of generalized metric and provides the orthogonal
splitting of the underlying vector space. In fact, if $\mathfrak{k}$ is of
coboundary type with quasitriangular factorizable $r$-matrix defining the
Lie bracket on $\mathfrak{k}^{\ast }$ through the cobracket in $\mathfrak{k}$%
, $\delta _{r}:\mathfrak{k}\longrightarrow \mathfrak{k}\otimes \mathfrak{k}$%
, as 
\begin{equation*}
\left\langle \left[ \xi ,\lambda \right] _{r},X\right\rangle =\left\langle
\xi \otimes \lambda ,\delta _{r}X\right\rangle
\end{equation*}%
then%
\begin{equation*}
\left[ \xi ,\lambda \right] _{r}=ad_{\hat{r}^{-}\left( \lambda \right)
}^{\ast }\xi -ad_{\hat{r}^{-}\left( \xi \right) }^{\ast }\lambda ,
\end{equation*}%
and the classical double of $\mathfrak{k}$, namely $\mathfrak{k}\oplus 
\mathfrak{k}^{\ast }$, is defined by the Lie bracket%
\begin{equation*}
\left[ X+\xi ,Y+\lambda \right] =\left[ X,Y\right] -ad_{\xi }^{r\ast
}Y+ad_{\lambda }^{r\ast }X+\left[ \xi ,\lambda \right] _{r}-ad_{X}^{\ast
}\lambda +ad_{Y}^{\ast }\xi
\end{equation*}%
As twilled extension, it corresponds to $\sigma _{X}\lambda =-ad_{X}^{\ast
}\lambda $ and $\rho _{\xi }Y=-ad_{\xi }^{r\ast }Y$, where $ad^{r\ast }$
means the transpose of the adjoint action by the Lie structure $\left[ ,%
\right] _{r}$ on $\mathfrak{k}^{\ast }$. Also we have that 
\begin{equation*}
ad_{\lambda }^{r\ast }X=\hat{r}^{-}\left( ad_{X}^{\ast }\lambda \right)
+ad_{X}\hat{r}^{-}\left( \lambda \right)
\end{equation*}%
that has the form of the $1$-cocycle introduced en $\left( \ref{tw 09}%
\right) $, so it is an exact twilled extension. Then the Lie bracket in $%
\mathfrak{k}\oplus \mathfrak{k}^{\ast }$ turns in 
\begin{eqnarray}
\left[ X+\xi ,Y+\lambda \right] &=&\left[ X,Y\right] +\left[ \hat{r}^{-}\xi
,Y\right] -\hat{r}^{-}ad_{Y}^{\ast }\xi -\left[ \hat{r}^{-}\lambda ,X\right]
+\hat{r}^{-}ad_{X}^{\ast }\lambda  \notag \\
&&+\left[ \xi ,\lambda \right] _{r}-ad_{X}^{\ast }\lambda +ad_{Y}^{\ast }\xi
.  \label{double 01}
\end{eqnarray}%
Note that together with the natural bilinear form 
\begin{equation*}
\left\langle \left( X,\xi \right) ,\left( Y,\lambda \right) \right\rangle
=\left\langle \xi ,Y\right\rangle +\left\langle X,\lambda \right\rangle ,
\end{equation*}%
$\left( \mathfrak{k}\oplus \mathfrak{k}^{\ast },\mathfrak{k},\mathfrak{k}%
^{\ast }\right) $ is a Manin triple.

On the other side, the twilled extension of $\mathfrak{k}$ and $\left( 
\mathfrak{k}^{\ast }\right) ^{op}$ is defined by the Lie bracket 
\begin{eqnarray*}
\left[ X+\xi ,Y+\lambda \right] ^{\prime } &=&\left[ X,Y\right] -\left[ \hat{%
r}^{-}\xi ,Y\right] +\hat{r}^{-}ad_{Y}^{\ast }\xi +\left[ \hat{r}^{-}\lambda
,X\right] -\hat{r}^{-}ad_{X}^{\ast }\lambda \\
&&-\left[ \xi ,\lambda \right] _{r}-ad_{X}^{\ast }\lambda +ad_{Y}^{\ast }\xi
\end{eqnarray*}%
The orthogonal subspaces $\mathcal{E}^{+},\mathcal{E}^{-}$ of decomposition $%
\left( \ref{cpd 04c}\right) $ are defined as 
\begin{equation*}
\mathcal{E}^{\pm }=\left\{ X+\left( \hat{r}^{-}\pm \hat{r}^{+}\right) \xi
/\xi \in \mathfrak{k}^{\ast }\right\}
\end{equation*}%
and the crossed Lie bracket is%
\begin{equation*}
\left[ \left( \xi ,\left( \hat{r}^{-}\pm \hat{r}^{+}\right) \xi \right)
,\left( \lambda ,\left( \hat{r}^{-}\pm \hat{r}^{+}\right) \lambda \right) %
\right] ^{\prime }=\pm 2\left( ad_{\hat{r}^{+}\lambda }^{\ast }\xi ,\left( 
\hat{r}^{-}\pm \hat{r}^{+}\right) ad_{\hat{r}^{+}\lambda }^{\ast }\xi
\right) .
\end{equation*}%
while%
\begin{equation*}
\left[ \left( \xi ,\left( \hat{r}^{-}+\hat{r}^{+}\right) \xi \right) ,\left(
\lambda ,\left( \hat{r}^{-}-\hat{r}^{+}\right) \lambda \right) \right] =0.
\end{equation*}%
Next we apply these results to a concrete example.

\subsubsection{\label{ex. sl2 1}Example: $\mathfrak{sl}_{2}$}

Let us consider the Lie algebra $\mathfrak{sl}_{2}$ spanned by the basis $%
\left\{ H,X_{+},X_{-}\right\} $ with the Lie brackets (Example 8.1.10 in 
\cite{Majid-book}) 
\begin{equation*}
\begin{array}{lllll}
\left[ H,X_{+}\right] =2X_{+} & , & \left[ H,X_{-}\right] =-2X_{-} & , & 
\left[ X_{+},X_{-}\right] =H%
\end{array}%
.
\end{equation*}%
Here we have the quasitriangular factorizable solution of the classical
Yang-Baxter equation $r\in \mathfrak{sl}_{2}\otimes \mathfrak{sl}_{2}$%
\begin{equation}
r=X_{+}\otimes X_{-}+\frac{1}{4}H\otimes H  \label{sl2 01}
\end{equation}%
whose symmetric and skew-symmetric parts are 
\begin{eqnarray*}
r_{+} &=&\frac{1}{2}X_{+}\otimes X_{-}+\frac{1}{2}X_{-}\otimes X_{+}+\frac{1%
}{4}H\otimes H \\
&& \\
r_{-} &=&\frac{1}{2}X_{+}\otimes X_{-}-\frac{1}{2}X_{-}\otimes X_{+}
\end{eqnarray*}%
The Lie cobracket $\delta _{r}:\mathfrak{sl}_{2}\longrightarrow \mathfrak{sl}%
_{2}\otimes \mathfrak{sl}_{2}$ is $\delta Z=ad_{Z}r=ad_{Z}r_{-}$, giving 
\begin{equation*}
\begin{array}{l}
\delta H=0 \\ 
\\ 
\delta X_{+}=\frac{1}{2}\left( X_{+}\otimes H-H\otimes X_{+}\right) \\ 
\\ 
\delta X_{-}=\frac{1}{2}\left( X_{-}\otimes H-H\otimes X_{-}\right)%
\end{array}%
.
\end{equation*}

Let $\mathfrak{sl}_{2}^{\ast }$ be the dual vector space of $\mathfrak{sl}%
_{2}$, then it turns into a Lie algebra with the Lie bracket defined as%
\begin{equation*}
\left\langle \left[ \eta ,\xi \right] _{r},Z\right\rangle =\left\langle \eta
\otimes \xi ,\delta _{r}Z\right\rangle
\end{equation*}%
therefore, if $\left\{ h,x_{+},x_{-}\right\} \subset \mathfrak{sl}_{2}^{\ast
}$ is the dual basis, we have that 
\begin{equation}
\begin{array}{lllll}
\left[ h,x_{+}\right] _{r}=-\frac{1}{2}x_{+} & , & \left[ h,x_{-}\right]
_{r}=-\frac{1}{2}x_{-} & , & \left[ x_{+},x_{-}\right] _{r}=0%
\end{array}%
.  \label{sl2 02}
\end{equation}%
This Lie algebra $\mathfrak{sl}_{2}^{\ast }$ is also a bialgebra with the
cobracket $\delta :\mathfrak{sl}_{2}^{\ast }\longrightarrow \mathfrak{sl}%
_{2}^{\ast }\otimes \mathfrak{sl}_{2}^{\ast }$ giving rise to the Lie
algebra in $\mathfrak{sl}_{2}$, then 
\begin{equation*}
\begin{array}{l}
\delta h=x_{+}\otimes x_{-}-x_{-}\otimes x_{+} \\ 
\\ 
\delta x_{+}=2\left( h\otimes x_{+}-x_{+}\otimes h\right) \\ 
\\ 
\delta x_{-}=2\left( x_{-}\otimes h-h\otimes x_{-}\right)%
\end{array}%
\end{equation*}

A nondegenerate symmetric bilinear form on $\mathfrak{sl}_{2}\oplus 
\mathfrak{sl}_{2}^{\ast }$, making the summands maximally isotropic
subalgebras, is defined by the pairing 
\begin{equation*}
\begin{array}{lllll}
\left\langle H,h\right\rangle =1 & , & \left\langle X_{+},x_{+}\right\rangle
=1 & , & \left\langle X_{-},x_{-}\right\rangle =1%
\end{array}%
\end{equation*}%
with all the other bracket vanishing.

The twilled extension giving rise to the classical double Lie algebra on $%
\mathfrak{sl}_{2}\oplus \mathfrak{sl}_{2}^{\ast }$ is defined by the Lie
bracket$\ \left( \ref{double 01}\right) $, which we write in terms of the
cobracket $\delta _{r}$ for ease of calculation 
\begin{eqnarray*}
\left[ \left( X,\xi \right) ,\left( Y,\lambda \right) \right] &=&\left[ X,Y%
\right] -\left( \xi \otimes I\right) \delta Y+\left( \lambda \otimes
I\right) \delta X \\
&&+\left[ \xi ,\lambda \right] _{r}-\left( X\otimes I\right) \delta \lambda
+\left( Y\otimes I\right) \delta \xi
\end{eqnarray*}%
From here we get the nonvanishing crossed Lie brackets

\begin{equation*}
\begin{array}{ll}
[X_{+},h]=-\frac{1}{2}X_{+}-x_{-}, & [X_{-},h]=-\frac{1}{2}X_{-}+x_{+} \\ 
&  \\ 
\lbrack H,x_{+}]=-2x_{+}, & [X_{+},x_{+}]=\frac{1}{2}H+2h \\ 
&  \\ 
\lbrack H,x_{-}]=2x_{-}, & [X_{-},x_{-}]=\frac{1}{2}H-2h%
\end{array}%
\end{equation*}

Let us now to build the generalized metric $\mathcal{G}\pm \mathcal{B}:%
\mathfrak{sl}_{2}^{\ast }\longrightarrow \mathfrak{sl}_{2}$ on $\mathfrak{sl}%
_{2}^{\ast }$ with%
\begin{equation*}
\begin{array}{ccc}
\mathcal{G}\xi =\left\langle \xi \otimes Id,r_{+}\right\rangle & \text{and}
& \mathcal{B}\xi =\left\langle \xi \otimes Id,r_{-}\right\rangle%
\end{array}%
\end{equation*}%
Thus we get%
\begin{equation*}
\begin{array}{ccccc}
\mathcal{G}h=\frac{1}{4}H & , & \mathcal{G}x_{+}=\frac{1}{2}X_{-} & , & 
\mathcal{G}x_{-}=\frac{1}{2}X_{+}%
\end{array}%
.
\end{equation*}%
The skew-symmetric part $\mathcal{B}$ is 
\begin{equation*}
\begin{array}{lllll}
\mathcal{B}h=0 & , & \mathcal{B}x_{+}=\frac{1}{2}X_{-} & , & \mathcal{B}%
x_{-}=-\frac{1}{2}X_{+}%
\end{array}%
.
\end{equation*}%
From these results we obtain $\left( \mathcal{B}\pm \mathcal{G}\right) $,
which has nontrivial unidimensional kernel. Of course, it verifies that%
\begin{equation*}
\left( \mathcal{B}\pm \mathcal{G}\right) \left[ x,y\right] _{\mathcal{B}}=%
\left[ \left( \mathcal{B}\pm \mathcal{G}\right) x,\left( \mathcal{B}\pm 
\mathcal{G}\right) y\right] .
\end{equation*}%
or, equivalently, one may check that $\mathcal{B}$ is an $\mathcal{O}$%
-operator of extension $\mathcal{G}$ with mass $-1$ 
\begin{equation*}
\lbrack \mathcal{B}x,\mathcal{B}y]-\mathcal{B}\Pi _{\mathfrak{sl}_{2}^{\ast
}}\left[ \mathcal{B}x,y\right] -\mathcal{B}\Pi _{\mathfrak{sl}_{2}^{\ast }}%
\left[ x,\mathcal{B}y\right] =-[\mathcal{G}x,\mathcal{G}y].
\end{equation*}%
The Lie algebra structure defined by $\mathcal{B}$, namely $\left( \left( 
\mathfrak{sl}_{2}^{\ast }\right) _{\mathcal{B}},\left[ ,\right] _{\mathcal{B}%
}\right) $, with 
\begin{equation*}
\left[ x,y\right] _{\mathcal{B}}=\Pi _{\mathfrak{sl}_{2}^{\ast }}\left[ 
\mathcal{B}x,y\right] -\Pi _{\mathfrak{sl}_{2}^{\ast }}\left[ \mathcal{B}y,x%
\right]
\end{equation*}%
gives rise to the Lie brackets $\left( \ref{sl2 02}\right) $.

The second double Lie algebra constructed as the\emph{\ twilled} \emph{%
extension} of $\mathfrak{sl}_{2}$ and $\left( \mathfrak{sl}_{2}^{\ast
}\right) ^{op}$, namely $\left( \mathfrak{sl}_{2}\oplus \left( \mathfrak{sl}%
_{2}^{\ast }\right) ^{op}\right) _{\mathcal{B}}$, has the Lie bracket%
\begin{eqnarray*}
&&\left[ \left( x,X\right) ,\left( y,Y\right) \right] ^{\prime } \\
&=&-\left[ x,y\right] _{\mathcal{B}}+\Pi _{\mathfrak{sl}_{2}^{\ast }}\left[
X,y\right] -\Pi _{\mathfrak{sl}_{2}^{\ast }}\left[ Y,x\right] \\
&&+\left[ X,Y\right] +\mathcal{B}\Pi _{\mathfrak{sl}_{2}^{\ast }}\left[ X,y%
\right] +\left[ \mathcal{B}y,X\right] -\mathcal{B}\Pi _{\mathfrak{sl}%
_{2}^{\ast }}\left[ Y,x\right] -\left[ \mathcal{B}x,Y\right]
\end{eqnarray*}%
that gives rise to the nonvanishing crossed Lie brackets

\begin{equation*}
\begin{array}{lll}
\left[ X_{+},h\right] ^{\prime }=\frac{1}{2}X_{+}-x_{-}, & [X_{-},h]^{\prime
}=x_{+}+\frac{1}{2}X_{-} &  \\ 
&  &  \\ 
\left[ H,x_{+}\right] ^{\prime }=-2x_{+}, & \left[ X_{+},x_{+}\right]
^{\prime }=2h-\frac{1}{2}H &  \\ 
&  &  \\ 
\lbrack H,x_{-}]^{\prime }=2x_{-}, & [X_{-},x_{-}]^{\prime }=-2h-\frac{1}{2}H
& 
\end{array}%
.
\end{equation*}

The orthogonal subspaces $\mathcal{E}^{+}$and $\mathcal{E}^{-}$ are spanned
by the graph of $\left( \mathcal{B}\pm \mathcal{G}\right) $ on the basis $%
\left\{ h,x_{+},x_{-}\right\} \subset \mathfrak{sl}_{2}^{\ast }$, so that 
\begin{eqnarray*}
\mathcal{E}^{+} &=&\overline{\left\{ h+\frac{1}{4}H,x_{+}+X_{-},x_{-}\right%
\} } \\
&& \\
\mathcal{E}^{-} &=&\overline{\left\{ h-\frac{1}{4}H,x_{+},x_{-}-X_{+}\right%
\} }
\end{eqnarray*}%
where the overbar is meant to indicate the linear span by each set of
vectors. The fundamental Lie brackets in each subspace are 
\begin{eqnarray*}
\left[ h+\frac{1}{4}H,x_{+}+X_{-}\right] ^{\prime } &=&-\left(
x_{+}+X_{-}\right)  \notag \\
\left[ h+\frac{1}{4}H,x_{-}\right] ^{\prime } &=&x_{-} \\
\left[ x_{+}+X_{-},x_{-}\right] ^{\prime } &=&-2\left( h+\frac{1}{4}H\right)
\notag
\end{eqnarray*}%
and%
\begin{eqnarray*}
\left[ h-\frac{1}{4}H,x_{+}\right] ^{\prime } &=&x_{+}  \notag \\
\left[ h-\frac{1}{4}H,x_{-}-X_{+}\right] ^{\prime } &=&-\left(
x_{-}-X_{+}\right) \\
\left[ x_{+},x_{-}-X_{+}\right] ^{\prime } &=&2\left( h-\frac{1}{4}H\right) 
\notag
\end{eqnarray*}%
showing that $\mathcal{E}^{+}$ and $\mathcal{E}^{-}$ are Lie subalgebras of $%
\left( \mathfrak{sl}_{2}^{\ast }\right) ^{op}\oplus \mathfrak{sl}_{2}$,
isomorphic to $\mathfrak{sl}_{2}$ and $\left( \mathfrak{sl}_{2}\right) ^{op}$%
, respectively. Of course, the crossed brackets between $\mathcal{E}^{+}$
and $\mathcal{E}^{-}$ vanish, confirming that they are ideals.

In the ordered basis $\left\{ h,x_{+},x_{-};H,X_{+},X_{-}\right\} \subset 
\mathfrak{sl}_{2}^{\ast }\oplus \mathfrak{sl}_{2}$, the linear operators%
\begin{equation*}
\begin{array}{ccc}
\mathcal{E}=\left( 
\begin{array}{cc}
-\mathcal{G}^{-1}\mathcal{B} & \mathcal{G}^{-1} \\ 
\mathcal{G}-\mathcal{BG}^{-1}\mathcal{B} & \mathcal{BG}^{-1}%
\end{array}%
\right)  & , & \mathcal{J}=\left( 
\begin{array}{cc}
-\mathcal{G}^{-1}\mathcal{B} & \mathcal{G}^{-1} \\ 
\mathcal{G}-\mathcal{BG}^{-1}\mathcal{B} & \mathcal{BG}^{-1}%
\end{array}%
\right) 
\end{array}%
\end{equation*}%
have associated the $6\times 6$ matrices%
\begin{equation*}
\begin{array}{ccc}
\mathcal{E}=\left( 
\begin{array}{cccccc}
0 & 0 & 0 & 4 & 0 & 0 \\ 
0 & 1 & 0 & 0 & 0 & 2 \\ 
0 & 0 & -1 & 0 & 2 & 0 \\ 
1/4 & 0 & 0 & 0 & 0 & 0 \\ 
0 & 0 & 0 & 0 & 1 & 0 \\ 
0 & 0 & 0 & 0 & 0 & -1%
\end{array}%
\right)  & , & \mathcal{J}=\left( 
\begin{array}{cccccc}
0 & 0 & 0 & 4 & 0 & 0 \\ 
0 & 1 & 0 & 0 & 0 & 2 \\ 
0 & 0 & -1 & 0 & 2 & 0 \\ 
-1/4 & 0 & 0 & 0 & 0 & 0 \\ 
0 & 0 & -1 & 0 & 1 & 0 \\ 
0 & -1 & 0 & 0 & 0 & -1%
\end{array}%
\right) 
\end{array}%
.
\end{equation*}

\bigskip

\section{From Lie algebras orthogonal direct sum to Manin triples}

In this section we will reverse the path followed in the previous section:
departing from a Lie algebra direct sum with a split bilinear form making
both the Lie subalgebras orthogonal, we construct a Manin triple containing
them as Lie ideals.

\subsection{Lie algebras direct sum and Manin quasi-triples}

Let $\mathfrak{g\ }$be the quadratic Lie algebra direct sum of the $n$%
-dimensional Lie algebras $\mathrm{E}^{+}$ and $\mathrm{E}^{-}$, with a
split bilinear form $\left( ,\right) _{\mathfrak{g}}$ rendering $\mathrm{E}%
^{+}$ and $\mathrm{E}^{-}$ mutually orthogonal. Note that Lie bracket of $%
\mathfrak{g}$ is 
\begin{equation*}
\left[ X^{+}+X^{-},Y^{+}+Y^{-}\right] _{\mathfrak{g}}=\left[ X^{+},Y^{+}%
\right] _{\mathrm{E}^{+}}+\left[ X^{-},Y^{-}\right] _{\mathrm{E}^{-}}
\end{equation*}%
implying that $\mathrm{E}^{+},\mathrm{E}^{-}$ are Lie ideals in $\mathfrak{g}
$.

We also assume that an antiisomorphism $\varphi :\mathrm{E}%
^{+}\longrightarrow \mathrm{E}^{-}$ such that $\varphi ^{\top }=-\varphi
^{-1}$ is given. As we saw in section \ref{cp vector spaces}, there exist a
couple of symmetric linear operators $\mathcal{E},\mathcal{J}:\mathfrak{g}%
\longrightarrow \mathfrak{g}$ with 
\begin{equation*}
\begin{array}{lllll}
\mathcal{E}^{2}=\mathcal{I} & , & \mathcal{J}^{2}=-\mathcal{I} & , & 
\mathcal{EJ}+\mathcal{JE}=0%
\end{array}%
\end{equation*}%
such that $\mathrm{E}^{+},\mathrm{E}^{-}$ are the $n$-dimensional orthogonal
eigenspaces of the product structure $\mathcal{E}$. In the direct sum
decomposition $\mathfrak{g}=\mathrm{E}^{+}\oplus \mathrm{E}^{-}$, the
block-matrices representing $\mathcal{E}$ and $\mathcal{J}$ are that given
in eq. $\left( \ref{cpd 05b}\right) $.

The eigenspaces of $\mathcal{F}=\mathcal{JE}$, namely $\mathrm{F}_{+}$ and $%
\mathrm{F}_{-}$ with%
\begin{equation*}
\mathrm{F}_{\pm }=\left\{ X^{+}\pm \varphi \left( X^{+}\right) /X^{+}\in 
\mathrm{E}^{+}\right\} ,
\end{equation*}%
give rise to the Lagrangian splitting $\mathfrak{g}=\mathrm{F}_{+}\oplus 
\mathrm{F}_{-}$, with the Lie algebra brackets 
\begin{eqnarray}
\left[ X^{+}+\varphi \left( X^{+}\right) ,Y^{+}-\varphi \left( Y^{+}\right) %
\right] &=&\left[ X^{+},Y^{+}\right] +\varphi \left( \left[ X^{+},Y^{+}%
\right] \right)  \notag \\
&&  \label{ybo 17f} \\
\left[ X^{+}\pm \varphi \left( X^{+}\right) ,Y^{+}\pm \varphi \left(
Y^{+}\right) \right] &=&\left[ X^{+},Y^{+}\right] -\varphi \left( \left[
X^{+},Y^{+}\right] \right)  \notag
\end{eqnarray}%
turning $\left( \mathfrak{g},\mathrm{F}_{+},\mathrm{F}_{-}\right) $ in a 
\emph{Manin quasi-triple } \cite{Alek-Kosm}, where $\mathrm{F}_{-}$ is a Lie
subalgebra while $\mathrm{F}_{+}$ is a $\mathrm{F}_{-}$-module,%
\begin{equation*}
\begin{array}{lllll}
\left[ \mathrm{F}_{-},\mathrm{F}_{-}\right] \subset \mathrm{F}_{-} & , & 
\left[ \mathrm{F}_{+},\mathrm{F}_{-}\right] \subset \mathrm{F}_{+} & , & 
\left[ \mathrm{F}_{+},\mathrm{F}_{+}\right] \subset \mathrm{F}_{-}%
\end{array}%
.
\end{equation*}%
In the case where $\mathfrak{g}$ is a semisimple Lie algebra, the pair $%
\left( \mathrm{F}_{-},\mathrm{F}_{+}\right) $ is a Cartan decomposition of $%
\mathfrak{g}$.

\subsection{Generalized metrics and $\mathcal{O}$-operators}

Consider now the linear bijection $\mathcal{G}:\mathrm{F}_{+}\longrightarrow 
\mathrm{F}_{-}$ defined in $\left( \ref{cpd 05a}\right) $, and we denote $%
X_{+}=X^{+}+\varphi \left( X^{+}\right) \in \mathrm{F}_{+}$. The original
orthogonal subspaces can be expressed as 
\begin{equation*}
\mathrm{E}^{\pm }=\left\{ X_{+}\pm \mathcal{G}\left( X_{+}\right)
/X_{+}=X^{+}+\varphi \left( X^{+}\right) \in \mathrm{F}_{+}\right\} .
\end{equation*}%
Therefore, in the Lagrangian splitting $\mathfrak{g}=\mathrm{F}_{+}\oplus 
\mathrm{F}_{-}$, the operators $\mathcal{E},\mathcal{J},\mathcal{F}$ are
represented by the block matrices%
\begin{equation*}
\mathcal{E}=\left( 
\begin{array}{cc}
0 & \mathcal{G}^{-1} \\ 
\mathcal{G} & 0%
\end{array}%
\right) \,\,\,\,,\,\,\,\,\mathcal{J}=\left( 
\begin{array}{cc}
0 & -\mathcal{G}^{-1} \\ 
\mathcal{G} & 0%
\end{array}%
\right) \,\,\,\,,\,\,\,\,\mathcal{F}=\left( 
\begin{array}{cc}
I & 0 \\ 
0 & -I%
\end{array}%
\right) .
\end{equation*}

In section \ref{cp vector spaces}, we saw that a wide family of orthogonal
subspaces is obtained by applying the gauge transformation $\left( \ref{cpd
06c}\right) $. They are parametrized by $\mathcal{B}\in \mathrm{Skew}\left( 
\mathrm{F}_{+},\mathrm{F}_{-}\right) $, namely 
\begin{equation*}
\mathrm{E}_{\mathcal{B}}^{\pm }=\mathrm{graph}\left( \mathcal{B}\pm \mathcal{%
G}\right) =\left\{ X_{+}+\left( \mathcal{B}\pm \mathcal{G}\right)
X_{+}/X_{+}\in \mathrm{F}_{+}\right\} ,
\end{equation*}%
such that $\mathfrak{g}=\mathrm{E}_{\mathcal{B}}^{+}\oplus \mathrm{E}_{%
\mathcal{B}}^{-}$. There is an complex product structure $\left( \mathcal{E}%
_{\mathcal{B}},\mathcal{J}_{\mathcal{B}}\right) $ associated with these
orthogonal subspaces that, in the direct sum decomposition $\mathfrak{g}=%
\mathrm{F}_{+}\oplus \mathrm{F}_{-}$, are 
\begin{equation*}
\begin{array}{lll}
\mathcal{E}_{\mathcal{B}}=\left( 
\begin{array}{cc}
-\mathcal{G}^{-1}\mathcal{B} & \mathcal{G}^{-1} \\ 
\mathcal{G}-\mathcal{BG}^{-1}\mathcal{B} & \mathcal{BG}^{-1}%
\end{array}%
\right) & , & \mathcal{J}_{\mathcal{B}}=\left( 
\begin{array}{cc}
-\mathcal{G}^{-1}\mathcal{B} & \mathcal{G}^{-1} \\ 
-\mathcal{G}-\mathcal{BG}^{-1}\mathcal{B} & \mathcal{BG}^{-1}%
\end{array}%
\right)%
\end{array}%
.
\end{equation*}

Following analogous steps as in subsection \ref{B as an O-operator}, we
intend to introduce a new Lie algebra structure in such a way that $\mathrm{E%
}_{\mathcal{B}}^{+}\ $and $\mathrm{E}_{\mathcal{B}}^{-}$ become Lie
subalgebras. This is attained by asking for $\mathcal{B}$ to be an $\mathcal{%
O}$-operator with extension $\mathcal{G}$ of mass $-1$ and considering the
vector subspace $\mathrm{F}_{+}$ as representation space of $\mathrm{F}_{-}$
under the bilinear map $\sigma :\mathrm{F}_{-}\longrightarrow \mathrm{End}%
\left( \mathrm{F}_{+}\right) $ defined as%
\begin{equation}
\sigma _{X_{-}}Y_{+}=\left[ X_{-},Y_{+}\right] .  \label{ybo 17f2}
\end{equation}

\begin{description}
\item[Proposition] $\mathcal{G}:\mathrm{F}_{+}\longrightarrow \mathrm{F}_{-} 
$ \textit{is} $\mathrm{F}_{-}$-\textit{invariant and antisymmetric, namely} 
\begin{equation*}
\begin{array}{lll}
\mathcal{G}\left( \sigma _{X_{-}}Y_{+}\right) =\left[ X_{-},\mathcal{G}Y_{+}%
\right] & , & \sigma _{\mathcal{G}X_{+}}Y_{+}+\sigma _{\mathcal{G}%
Y_{+}}X_{+}=0%
\end{array}%
.
\end{equation*}
\end{description}

\textbf{Proof}: Recall that $\mathcal{G}:\mathrm{F}_{+}\longrightarrow 
\mathrm{F}_{-}$ is defined as in $\left( \ref{cpd 05a}\right) $, then using
the Lie brackets $\left( \ref{ybo 17f}\right) $, it is immediate to see that 
\begin{equation*}
\mathcal{G}\left( \sigma _{X_{-}}Y_{+}\right) =\left[ X_{-},\mathcal{G}Y_{+}%
\right] .
\end{equation*}%
For the antisymmetry consider 
\begin{equation*}
\mathcal{G}\left( \sigma _{\mathcal{G}X_{+}}Y_{+}+\sigma _{\mathcal{G}%
Y_{+}}X_{+}\right) =\mathcal{G}\left( \left[ \mathcal{G}X_{+},Y_{+}\right] +%
\left[ \mathcal{G}Y_{+},X_{+}\right] \right)
\end{equation*}%
and, by the previous result, the r.h.s. vanish.$\blacksquare $

Next we seek a linear operator $\mathcal{B}:\mathrm{F}_{+}\longrightarrow 
\mathrm{F}_{-}$ fulfilling the equation. 
\begin{equation*}
\left[ \mathcal{B}X_{+},\mathcal{B}Y_{+}\right] -\mathcal{B}\left( \sigma _{%
\mathcal{B}X_{+}}Y_{+}-\sigma _{\mathcal{B}Y_{+}}X_{+}\right) =-\left[ 
\mathcal{G}X_{+},\mathcal{G}Y_{+}\right] ,
\end{equation*}%
so $\mathcal{B}$ is an $\mathcal{O}$-operator with extension $\mathcal{G}$
of mass $-1$. In turn, this implies that

\begin{enumerate}
\item \textit{The bracket}%
\begin{equation*}
\left[ X_{+},Y_{+}\right] _{\mathcal{B}}=\left[ \mathcal{B}X_{+},Y_{+}\right]
-\left[ \mathcal{B}Y_{+},X_{+}\right]
\end{equation*}%
\textit{defines a Lie algebra on} $\mathrm{F}_{+}$,

\item $\left( \mathcal{B}\pm \mathcal{G}\right) :\mathrm{F}%
_{+}\longrightarrow \mathrm{F}_{-}$ \textit{is a Lie algebra homomorphism
from }$\left( \mathrm{F}_{+},\left[ ,\right] _{\mathcal{B}}\right) $ \textit{%
to} $\left( \mathrm{F}_{-},\left[ ,\right] \right) $\textit{, namely} 
\textit{\ }%
\begin{equation*}
\left( \mathcal{B}\pm \mathcal{G}\right) \left[ X_{+},Y_{+}\right] _{%
\mathcal{B}}=\left[ \left( \mathcal{B}\pm \mathcal{G}\right) X_{+},\left( 
\mathcal{B}\pm \mathcal{G}\right) Y_{+}\right]
\end{equation*}%
$\forall X_{+},Y_{+}\in $ $\mathrm{F}_{+}$.
\end{enumerate}

Let us denote the Lie algebra $\left( \mathrm{F}_{+},\left[ ,\right] _{%
\mathcal{B}}\right) $ as $\mathrm{F}_{+}^{\mathcal{B}}$. There is a
one-to-one correspondence between solutions to the $\mathcal{O}$-operator
condition $\left( \ref{Bop 03b}\right) $ on $\mathrm{F}_{-}$ and solutions
of the modified classical Yang-Baxter equation on $\mathrm{E}^{+}$.
Recalling that $\mathrm{F}_{\pm }=\mathrm{graph}\left( \pm \varphi \right) $
we may think of $\mathcal{B}:\mathrm{F}_{+}\rightarrow \mathrm{F}_{-}$ as
realized by some linear operator $\theta :\mathrm{E}^{+}\longrightarrow 
\mathrm{E}^{+}$ in such a way that 
\begin{equation*}
\mathcal{B}\left( X^{+}+\varphi \left( X^{+}\right) \right) =\theta
X^{+}-\varphi \left( \theta X^{+}\right) .
\end{equation*}

\begin{description}
\item[Theorem] $\mathcal{B}:\mathrm{F}_{+}\rightarrow \mathrm{F}_{-}$ 
\textit{is an extended }$\mathcal{O}$\textit{-operator with extension} $%
\mathcal{G}$ \textit{of mass} $\kappa =-1$ \textit{if and only if }$\theta :%
\mathrm{E}^{+}\longrightarrow \mathrm{E}^{+}$ \textit{is a skew-symmetric
solution to the modified classical Yang-Baxter equation.}
\end{description}

\textbf{Proof: }From the eq. $\left( \ref{Bop 03b}\right) $ written in terms
of $X_{+}=X^{+}+\varphi \left( X^{+}\right) $ and $Y_{+}=Y^{+}+\varphi
\left( Y^{+}\right) $ and having in mind the explicit form of the left
action $\sigma _{\mathcal{B}X_{+}}Y_{+}$, 
\begin{equation*}
\sigma _{\left( \theta X^{+}-\varphi \left( \theta X^{+}\right) \right)
}\left( Y^{+}+\varphi \left( Y^{+}\right) \right) =\left[ \theta X^{+},Y^{+}%
\right] +\varphi \left[ \theta X^{+},Y^{+}\right]
\end{equation*}%
and the same for $\sigma _{\mathcal{B}Y_{+}}X_{+}$, one gets 
\begin{eqnarray*}
&&\left[ \theta X^{+},\theta Y^{+}\right] -\varphi \left( \left[ \theta
X^{+},\theta Y^{+}\right] \right) \\
&&-\mathcal{B}\left( \left( \left[ \theta X^{+},Y^{+}\right] ,\varphi \left[
\theta X^{+},Y^{+}\right] \right) \right) \\
&&-\mathcal{B}\left( \left( \left[ \theta Y^{+},X^{+}\right] ,\varphi \left[
\theta Y^{+},X^{+}\right] \right) \right) \\
&=&-\left[ \left( X^{+}-\varphi \left( X^{+}\right) \right) ,\left(
Y^{+}-\varphi \left( Y^{+}\right) \right) \right]
\end{eqnarray*}%
Applying again the action of $\mathcal{B}$ in terms of $\theta $, it turns in%
\begin{eqnarray*}
&&\left[ \theta X^{+},\theta Y^{+}\right] -\theta \left( \left[ \theta
X^{+},Y^{+}\right] -\left[ \theta Y^{+},X^{+}\right] \right) +\left[
X^{+},Y^{+}\right] \\
&=&\varphi \left( \left[ \theta X^{+},\theta Y^{+}\right] -\theta \left( %
\left[ \theta X^{+},Y^{+}\right] -\left[ \theta Y^{+},X^{+}\right] \right) +%
\left[ X^{+},Y^{+}\right] \right)
\end{eqnarray*}%
which holds if and only if 
\begin{equation*}
\left[ \theta X^{+},\theta Y^{+}\right] -\theta \left[ \theta X^{+},Y^{+}%
\right] -\theta \left[ X^{+},\theta Y^{+}\right] =-\left[ X^{+},Y^{+}\right]
\end{equation*}%
so $\theta :\mathrm{E}^{+}\longrightarrow \mathrm{E}^{+}$ is a solution of
the modified classical Yang-Baxter equation on $\mathrm{E}^{+}$.

On the other side, because $\mathcal{B}$ is skew-symmetric relative to the
bilinear form $\left( ,\right) _{\mathfrak{g}}$, $\theta $ must be also
skew-symmetric.$\blacksquare $

\subsection{Twilled extension of $\mathrm{F}_{+}^{\mathcal{B}},\mathrm{F}%
_{-} $}

So far, we have built a Lagrangian decomposition $\mathfrak{g}=\mathrm{F}%
_{+}\oplus \mathrm{F}_{-}$ of a Lie algebra direct sum $\mathfrak{g}=\mathrm{%
E}^{+}\oplus \mathrm{E}^{-}$, where $\mathrm{F}_{-}$ is a Lie subalgebra.
Then, by introducing the generalized metric $\mathcal{G}\pm \mathcal{B}$ and
regarding it as an $\mathcal{O}$-operator, we obtained a Lie algebra
structure on $\mathrm{F}_{+}$, denoted as $\mathrm{F}_{+}^{\mathcal{B}}$,
which is homomorphic to $\mathrm{F}_{-}$. Next, by constructing a \emph{%
twilled extension} out of the Lie algebras $\left( \mathrm{F}_{+}^{\mathcal{B%
}}\right) ^{op},\mathrm{F}_{-}$ \cite{Kosmann-Magri-01}, we shall get a new
Lie algebra structure on $\mathfrak{g}\ $having $\left( \mathrm{F}_{+}^{%
\mathcal{B}}\right) ^{op}$ and $\mathrm{F}_{-}$ as Lie subalgebras.

Together with map $\sigma :\mathrm{F}_{-}\longrightarrow \mathrm{End}\left( 
\mathrm{F}_{+}\right) $ given in eq. $\left( \ref{ybo 17f2}\right) $ and
following the results of subsection \ref{twilled extensions}, in particular
eq. $\left( \ref{tw 09}\right) $, we introduced the map\textit{\ }$\rho
:\left( \mathrm{F}_{+}^{\mathcal{B}}\right) ^{op}\longrightarrow \mathrm{End}%
\left( \mathrm{F}_{-}\right) $ defined as 
\begin{equation}
\rho _{X_{+}}Y_{-}=\mathcal{B}\left[ X_{+},Y_{-}\right] -\left[ \mathcal{B}%
X_{+},Y_{-}\right]  \label{LbgB 00}
\end{equation}%
It is a $1$-cocycle making the pair on maps $\sigma $,$\rho $ to fulfill the
constraints $\left( \ref{tw 04}\right) $, so we get a twilled extension $%
\mathfrak{g}_{\mathcal{B}}=\left( \mathrm{F}_{+}^{\mathcal{B}}\right)
^{op}\oplus \mathrm{F}_{-}$ of $\left( \mathrm{F}_{+}^{\mathcal{B}}\right)
^{op}$ and $\mathrm{F}_{-}$ with Lie bracket 
\begin{equation*}
\begin{array}{l}
\left[ X_{+}+X_{-},Y_{+}+Y_{-}\right] \\ 
=-\left[ X_{+},Y_{+}\right] _{\mathcal{B}}+\left[ X_{-},Y_{+}\right] -\left[
Y_{-},X_{+}\right] \\ 
~~~\oplus \left[ X_{-},Y_{-}\right] -\left[ \mathcal{B}X_{+},Y_{-}\right] -%
\mathcal{B}\left[ Y_{-},X_{+}\right] +\left[ \mathcal{B}Y_{+},X_{-}\right] +%
\mathcal{B}\left[ X_{-},Y_{+}\right] .%
\end{array}%
\end{equation*}

\begin{description}
\item[Proposition] $\left( \mathfrak{g}_{\mathcal{B}},\left( \mathrm{F}_{+}^{%
\mathcal{B}}\right) ^{op},\mathrm{F}_{-}\right) $ \textit{equipped with the
bilinear form} $\left( ,\right) _{\mathfrak{g}}$ \textit{is a Manin triple.}
\end{description}

\textbf{Proof: }It just remains to prove that the bilinear form $\left(
,\right) _{\mathfrak{g}}$ is invariant under the adjoint action of $%
\mathfrak{g}_{\mathcal{B}}$. Let $\left( X_{+}+X_{-}\right) $ , $\left(
Y_{+}+Y_{-}\right) $ , $\left( Z_{+}+Z_{-}\right) \in \mathfrak{g}_{\mathcal{%
B}}$ then

\begin{eqnarray*}
&&\left( \left[ X_{+}+X_{-},Y_{+}+Y_{-}\right] _{\mathfrak{g}_{\mathcal{B}%
}},Z_{+}+Z_{-}\right) _{\mathfrak{g}} \\
&=&\left( Y_{+},\left[ \mathcal{B}X_{+},Z_{-}\right] -\mathcal{B}\left[
X_{+},Z_{-}\right] -\left[ X_{-},Z_{-}\right] \right) _{\mathfrak{g}} \\
&&+\left( Y_{+}, \left[ X_{-},\mathcal{B}Z_{+}\right] +\mathcal{B}\left[
Z_{+},X_{-}\right] \right) _{\mathfrak{g}} \\
&&+\left( Y_{-},\left[ Z_{-},X_{+}\right] -\left[ X_{-},Z_{+}\right] -\left[ 
\mathcal{B}Z_{+},X_{+}\right] +\left[ \mathcal{B}X_{+},Z_{+}\right] \right)
_{\mathfrak{g}} \\
&=&-\left( Y_{-},-\left[ \mathcal{B}X_{+},Z_{+}\right] -\left[ X_{+},%
\mathcal{B}Z_{+},\right] +\left[ X_{-},Z_{+}\right] -\left[ Z_{-},X_{+}%
\right] \right) _{\mathfrak{g}} \\
&&-\left( Y_{+},\left[ X_{-},Z_{-}\right] +\mathcal{B}\left[ X_{-},Z_{+}%
\right] +\left[ \mathcal{B}Z_{+},X_{-}\right] \right) _{\mathfrak{g}} \\
&&+\left( Y_{+},\mathcal{B}\left[ Z_{-},X_{+}\right] +\left[ \mathcal{B}%
X_{+},Z_{-}\right] \right) _{\mathfrak{g}} \\
&=&-\left( \left( Y_{+},Y_{-}\right) ,\left[ \left( X_{+},X_{-}\right)
,\left( Z_{+},Z_{-}\right) \right] _{\mathfrak{g}_{\mathcal{B}}}\right) _{%
\mathfrak{g}}
\end{eqnarray*}%
as expected. Thus it shows that $\left( \mathfrak{g}_{\mathcal{B}},\left( 
\mathrm{F}_{+}^{\mathcal{B}}\right) ^{op},\mathrm{F}_{-}\right) $\textit{\ }%
with the nondegenerate invariant symmetric bilinear form $\left( ,\right) _{%
\mathfrak{g}}$ is a Manin triple\textit{.}$\blacksquare $

This Manin triple $\left( \mathfrak{g}_{\mathcal{B}},\left( \mathrm{F}_{+}^{%
\mathcal{B}}\right) ^{op},\mathrm{F}_{-}\right) $ also admit an orthogonal
splitting in Lie ideals $\mathfrak{g}_{\mathcal{B}}=\mathrm{E}_{\mathcal{B}%
}^{+}\oplus \mathrm{E}_{\mathcal{B}}^{-}$, with $\mathrm{E}_{\mathcal{B}%
}^{\pm }=\left\{ X_{+}+\left( \mathcal{B}\pm \mathcal{G}\right)
X_{+}/X_{+}\in \mathrm{F}_{+}\right\} $. The Lie brackets in these
subalgebras are 
\begin{eqnarray*}
&&\left[ \left( X_{+},\left( \mathcal{B}\pm \mathcal{G}\right) X_{+}\right)
,\left( Y_{+},\left( \mathcal{B}\pm \mathcal{G}\right) Y_{+}\right) \right]
_{\mathfrak{g}_{\mathcal{B}}} \\
&=&\pm 2\left( \mathcal{G}^{-1}\left[ \mathcal{G}X_{+},\mathcal{G}Y_{+}%
\right] +\left( \mathcal{B}\pm \mathcal{G}\right) \mathcal{G}^{-1}\left[ 
\mathcal{G}X_{+},\mathcal{G}Y_{+}\right] \right)
\end{eqnarray*}%
and 
\begin{equation*}
\left[ \left( X_{+},\left( \mathcal{B}+\mathcal{G}\right) X_{+}\right)
,\left( Y_{+},\left( \mathcal{B}-\mathcal{G}\right) Y_{+}\right) \right] _{%
\mathfrak{g}_{\mathcal{B}}}=0.
\end{equation*}

\subsection{Example: Anti-isomorphic Lie algebras}

Let $\mathrm{E}^{+}$ and $\mathrm{E}^{-}$ be a couple of Lie algebras
connected by an antiisomorphism $\varphi :\mathrm{E}^{+}\longrightarrow 
\mathrm{E}^{-}$, and $\mathrm{E}^{-}$ equipped with an invariant, symmetric
nondegenerate bilinear form $\left( ,\right) _{\mathrm{E}^{-}}$. Then a
symmetric nondegenerate bilinear form on $\mathrm{E}^{+}$ can be defined as 
\begin{equation*}
\left( X^{+},Y^{+}\right) _{\mathrm{E}^{+}}=-\left( \varphi \left(
X^{+}\right) ,\varphi \left( Y^{+}\right) \right) _{\mathrm{E}^{-}}
\end{equation*}%
which is also $\mathrm{E}^{+}$-invariant and implies $\varphi ^{\top
}=-\varphi ^{-1}$.

Consider the Lie algebra direct sum $\mathfrak{g}=\mathrm{E}^{+}\oplus 
\mathrm{E}^{-}$ equipped with the invariant non-degenerate symmetric
bilinear form $\left( ,\right) _{\mathfrak{g}}:\mathfrak{g}\otimes \mathfrak{%
g}\longrightarrow \mathbb{R}$ defined as 
\begin{equation*}
\left( \left( X^{+},X^{-}\right) ,\left( Y^{+},Y^{-}\right) \right) _{%
\mathfrak{g}}=\left( X^{+},Y^{+}\right) _{\mathrm{E}^{+}}+\left(
X^{-},Y^{-}\right) _{\mathrm{E}^{-}}
\end{equation*}%
making $\mathrm{E}^{+}\ $and $\mathrm{E}^{-}$ mutually orthogonal subspaces
of $\mathfrak{g}$.

The graph of $\pm \varphi $ provides a Lagrangian splitting $\mathfrak{g}=%
\mathrm{F}_{+}\oplus \mathrm{F}_{-}$ where%
\begin{equation*}
\mathrm{F}_{\pm }=\left\{ \left( X^{+},\pm \varphi \left( X^{+}\right)
\right) /X^{+}\in \mathrm{E}^{+}\right\} .
\end{equation*}%
so the invertible linear map $\mathcal{G}:\mathrm{F}_{+}\longrightarrow 
\mathrm{F}_{-}$ introduced in eq. $\left( \ref{cpd 05a}\right) $ is defined
as%
\begin{equation*}
\mathcal{G}\left( X^{+},\varphi \left( X^{+}\right) \right) =\left(
X^{+},-\varphi \left( X^{+}\right) \right)
\end{equation*}

For instance, consider the Lie algebra $\mathrm{E}^{+}=\mathfrak{h}$
equipped with a symmetric, nondegenerate, ad-invariant bilinear form $\left(
,\right) _{\mathfrak{h}}$, and $\mathrm{E}^{-}=\mathfrak{h}^{op}$, with the
antihomomorphism $\varphi :\mathfrak{h}\longrightarrow \mathfrak{h}^{op}$
defined as $\varphi \left( X\right) =X$ for $X\in \mathfrak{h}$, then $\left[
\varphi \left( X\right) ,\varphi \left( Y\right) \right] _{\mathfrak{h}%
^{op}}=-\left[ X,Y\right] =-\varphi \left( \left[ X,Y\right] \right) $.
Define the Lie algebra direct sum $\mathfrak{g}=\mathfrak{h}\oplus \mathfrak{%
h}^{op}$ with the Lie bracket%
\begin{equation*}
\left[ \left( X,X^{\prime }\right) ,\left( Y,Y^{\prime }\right) \right] _{%
\mathfrak{g}}=\left( \left[ X,Y\right] ,-\left[ X^{\prime },Y^{\prime }%
\right] \right)
\end{equation*}%
and the bilinear form 
\begin{equation*}
\left( \left( X,X^{\prime }\right) ,\left( Y,Y^{\prime }\right) \right) _{%
\mathfrak{g}}=\left( X,Y\right) _{\mathfrak{h}}-\left( X^{\prime },Y^{\prime
}\right) _{\mathfrak{h}}
\end{equation*}%
which is invariant and makes $\mathfrak{h}$ and $\mathfrak{h}^{op}$
orthogonal Lie subalgebras of $\mathfrak{g}$. The subspaces 
\begin{equation*}
\mathfrak{h}_{\pm }=\left\{ \left( X,\pm X\right) /X\in \mathfrak{h}\right\}
=\mathrm{graph}\left( \pm \varphi \right)
\end{equation*}%
are Lagrangian and $\mathfrak{g}=\mathfrak{h}_{+}\oplus \mathfrak{h}_{-}$,
and the Lie brackets in $\mathfrak{h}_{\pm }$ are%
\begin{eqnarray*}
\left[ \left( X,\pm X\right) ,\left( Y,\pm Y\right) \right] _{\mathfrak{g}}
&=&\left( \left[ X,Y\right] ,-\left[ X,Y\right] \right) \in \mathfrak{h}_{-}
\\
\left[ \left( X,+X\right) ,\left( Y,-Y\right) \right] _{\mathfrak{g}}
&=&\left( \left[ X,Y\right] ,+\left[ X,Y\right] \right) \in \mathfrak{h}_{+}
\end{eqnarray*}

Now, consider the linear map $\mathcal{G}:\mathfrak{h}_{+}\longrightarrow 
\mathfrak{h}_{-}$ is 
\begin{equation*}
\mathcal{G}\left( X,+X\right) =\left( X,-X\right)
\end{equation*}%
for $X\in \mathfrak{h}$, which is $\mathfrak{h}_{-}$-invariant:%
\begin{equation*}
\mathcal{G}\left( \left[ \left( X,-X\right) ,\left( Y,+Y\right) \right]
\right) =\left( \left[ X,Y\right] ,-\left[ X,Y\right] \right) =\left[ \left(
X,-X\right) ,\mathcal{G}\left( Y,+Y\right) \right]
\end{equation*}%
and antisymmetric. Together with $\mathcal{G}$, we introduce a linear
operator $\mathcal{B}:\mathfrak{h}_{+}\longrightarrow \mathfrak{h}_{-}$
satisfying the $\mathcal{O}$-operator condition $\left( \ref{Bop 02a}\right) 
$, which is realized by some linear operator $\theta :\mathfrak{h}%
\longrightarrow \mathfrak{h}$ such that 
\begin{equation*}
\mathcal{B}\left( X,+X\right) =\left( \theta X,-\theta X\right)
\end{equation*}%
provided $\theta $ is a solution of the modified classical Yang-Baxter
equation on $\mathfrak{h}$. With these operator we get a new Lie algebra
structure on $\mathfrak{h}_{+}$ with Lie bracket%
\begin{equation*}
\left[ \left( X,+X\right) ,\left( Y,+Y\right) \right] _{\mathcal{B}}=\left( %
\left[ X,Y\right] _{\theta },\left[ X,Y\right] _{\theta }\right) .
\end{equation*}%
where, for $X,Y\in \mathfrak{h}$ 
\begin{equation*}
\left[ X,Y\right] _{\theta }=\left[ \theta X,Y\right] -\left[ \theta Y,X%
\right]
\end{equation*}%
We denote $\left( \mathfrak{h}_{+},\left[ ,\right] _{\theta }\right) $ as $%
\mathfrak{h}_{+}^{\theta }$.

The next step is to construct the twilled action, so we need the map $\sigma
:\mathfrak{h}_{-}\longrightarrow \mathrm{End}\left( \mathfrak{h}_{+}\right) $
$\left( \ref{ybo 17f2}\right) $ that is defined here as%
\begin{equation*}
\sigma _{\left( X,-X\right) }\left( Y,+Y\right) =\left[ \left( X,-X\right)
,\left( Y,+Y\right) \right] =\left[ X,Y\right] +\left[ X,Y\right] \in 
\mathfrak{h}_{+}
\end{equation*}%
and the map $\rho :\mathfrak{h}_{+}\longrightarrow \mathrm{End}\left( 
\mathfrak{h}_{-}\right) $ $\left( \ref{LbgB 00}\right) $ which turns in 
\begin{equation*}
\rho _{\left( X,+X\right) }\left( Y,-Y\right) =\left( \theta \left[ X,Y%
\right] -\left[ \theta X,Y\right] ,-\theta \left[ X,Y\right] -\left[ \theta
X,Y\right] \right)
\end{equation*}%
Thus the twilled extension $\mathfrak{g}_{\mathcal{B}}=\left( \mathfrak{h}%
_{+}^{\theta }\right) ^{op}\oplus \mathfrak{h}_{-}$ of $\left( \mathfrak{h}%
_{+}^{\theta }\right) ^{op}$ and $\mathfrak{h}_{-}$ is defined by the Lie
bracket 
\begin{equation*}
\begin{array}{l}
\left[ \left( X,+X\right) +\left( X^{\prime },-X^{\prime }\right) ,\left(
Y,+Y\right) +\left( Y^{\prime },-Y^{\prime }\right) \right] \\ 
=-\left[ \left( X,+X\right) ,\left( Y,+Y\right) \right] _{\mathcal{B}}+\left[
\left( X^{\prime },-X^{\prime }\right) ,\left( Y,+Y\right) \right] \\ 
-\left[\left( Y^{\prime },-Y^{\prime }\right) ,\left( X,+X\right) \right] +%
\left[ \left( X^{\prime },-X^{\prime }\right) ,\left( Y^{\prime },-Y^{\prime
}\right) \right] \\ 
~~~ -\left[ \mathcal{B}\left( X,+X\right) ,\left( Y^{\prime },-Y^{\prime
}\right) \right] -\mathcal{B}\left[ \left( Y^{\prime },-Y^{\prime }\right)
,\left( X,+X\right) \right] \\ 
~~~+\left[ \mathcal{B}\left( Y,+Y\right) ,\left( X^{\prime },-X^{\prime
}\right) \right] +\mathcal{B}\left[ \left( X^{\prime },-X^{\prime }\right)
,\left( Y,+Y\right) \right].%
\end{array}%
\end{equation*}

The orthogonal splitting $\mathfrak{g}_{\theta }=\mathfrak{h}_{\theta
}^{+}\oplus \mathfrak{h}_{\theta }^{-}$, is defined by the graph of $\left( 
\mathcal{B}\pm \mathcal{G}\right) $ namely 
\begin{eqnarray*}
\mathfrak{h}_{\theta }^{+} &=&\left\{ \left( X,+X\right) +\left( \theta
X+X,-\left( \theta X+X\right) \right) /X\in \mathfrak{h}\right\} \\
\mathfrak{h}_{\theta }^{-} &=&\left\{ \left( X,+X\right) +\left( \theta
X-X,-\left( \theta X-X\right) \right) /X\in \mathfrak{h}\right\} .
\end{eqnarray*}%
Let us introduce the map $\Theta ^{\pm }:\mathfrak{h}\longrightarrow 
\mathfrak{h}_{\theta }^{\pm }$ such that $\Theta ^{\pm }\left( X\right)
=\left( X,+X\right) +\left( \theta X\pm X,-\left( \theta X+X\right) \right) $%
, then the Lie bracket in $\mathfrak{h}_{\theta }^{\pm }$ is 
\begin{equation*}
\left[ \Theta ^{\pm }\left( X\right) ,\Theta ^{\pm }\left( Y\right) \right]
=\pm 2\left( \Theta ^{\pm }\left( \left[ X,Y\right] \right) \right)
\end{equation*}%
showing that $\mathfrak{h}_{\theta }^{\pm }$ is a Lie subalgebra and that
the linear map $\frac{1}{2}\Theta ^{\pm }:\mathfrak{h}\longrightarrow 
\mathfrak{h}_{\theta }^{\pm }$ is a Lie algebra isomorphism. Of course, the
crossed bracket is 
\begin{equation*}
\left[ \Theta ^{+}\left( X\right) ,\Theta ^{-}\left( Y\right) \right]
=\left( 0,0\right)
\end{equation*}%
verifying that $\mathfrak{h}_{\theta }^{\pm }$ is an ideal.

\subsubsection{Example: $\mathfrak{sl}_{2}\left( \mathbb{C}\right) $}

The above results can be applied to the example of subsection \ref{ex. sl2 1}%
: Define 
\begin{equation*}
\begin{array}{lll}
a_{1}=2\left( h-\frac{1}{4}H\right) & , & b_{1}=2\left( h+\frac{1}{4}H\right)
\\ 
&  &  \\ 
a_{2}=x_{+} & , & b_{2}=\left( x_{+}+X_{-}\right) \\ 
&  &  \\ 
a_{3}=\left( x_{-}-X_{+}\right) & , & b_{3}=x_{-}%
\end{array}%
\end{equation*}%
then consider the Lie algebras $\mathfrak{h}^{-}=\mathfrak{sl}_{2}$ and $%
\mathfrak{h}^{+}=\left( \mathfrak{sl}_{2}\right) ^{op}$ generated by the
basis $\left\{ a_{1},a_{2},a_{3}\right\} $ and $\left\{
b_{1},b_{2},b_{3}\right\} $, respectively, with the Lie brackets%
\begin{equation*}
\begin{array}{ccc}
\left[ a_{1},a_{2}\right] =2a_{2} & \left[ a_{1},a_{3}\right] =-2a_{3} & 
\left[ a_{2},a_{3}\right] =a_{1}%
\end{array}%
\end{equation*}%
and%
\begin{equation*}
\begin{array}{ccc}
\left[ b_{1},b_{2}\right] =-2b_{2} & \left[ b_{1},b_{3}\right] =2b_{3} & 
\left[ b_{2},b_{3}\right] =-b_{1}%
\end{array}%
.
\end{equation*}%
Let $\mathfrak{h}^{-}$ be equipped with a symmetric nondegenerate invariant
bilinear form $\left( ,\right) _{\mathfrak{h}^{-}}$ defined by the matrix%
\begin{equation*}
\left( a_{i},a_{j}\right) _{\mathfrak{h}^{-}}=\left( 
\begin{array}{ccc}
2 & 0 & 0 \\ 
0 & 0 & 1 \\ 
0 & 1 & 0%
\end{array}%
\right)
\end{equation*}%
which is symmetric, nondegenerate and invariant. We use it to equip the Lie
algebra direct sum $\mathfrak{g}=\mathfrak{h}^{+}\oplus \mathfrak{h}^{-}$
with the bilinear form 
\begin{equation*}
\left( X^{+}+X^{-},Y^{+}+Y^{-}\right) _{\mathfrak{g}}=\left(
X^{-},Y^{-}\right) _{\mathfrak{h}^{-}}-\left( \varphi \left( X^{+}\right)
,\varphi \left( Y^{+}\right) \right) _{\mathfrak{h}^{-}}
\end{equation*}%
making $\mathfrak{h}^{+}$ and $\mathfrak{h}^{-}$ mutually orthogonal.

The linear map $\varphi :\mathfrak{h}^{+}\longrightarrow \mathfrak{h}^{-}$
defined as $\varphi \left( b_{i}\right) =a_{i}$ is an antiisomorphism of Lie
algebras and the graph of $\pm \varphi $ is 
\begin{equation*}
\mathfrak{h}_{\pm }=\mathrm{graph}\left( \pm \varphi \right) =\overline{%
\left\{ b_{1}\pm a_{1},b_{2}\pm a_{2},b_{3}\pm a_{3}\right\} }
\end{equation*}%
Then%
\begin{eqnarray*}
&&\left( b_{i}\pm \varphi \left( b_{i}\right) ,b_{j}\pm \varphi \left(
b_{j}\right) \right) _{\mathfrak{g}} \\
&&=\left( \varphi \left( b_{i}\right) ,\varphi \left( b_{j}\right) \right) _{%
\mathfrak{h}^{-}}-\left( \varphi \left( b_{i}\right) ,\varphi \left(
b_{j}\right) \right) _{\mathfrak{h}^{-}} \\
&&=0
\end{eqnarray*}%
so that $\mathfrak{h}_{\pm }$ is a Lagrangian subspace.

Let us denote $e_{\pm }^{i}=b_{i}\pm \varphi \left( b_{i}\right) =b_{i}\pm
a_{i}$, then the nonvanishing Lie brackets in the Lie algebra direct sum are%
\begin{equation*}
\begin{array}{lllll}
\left[ e_{+}^{1},e_{+}^{2}\right] =-2e_{-}^{2} & , & \left[
e_{+}^{1},e_{+}^{3}\right] =2e_{-}^{3} & , & \left[ e_{+}^{2},e_{+}^{3}%
\right] =-e_{-}^{1} \\ 
&  &  &  &  \\ 
\left[ e_{-}^{1},e_{-}^{2}\right] =-2e_{-}^{2} & , & \left[
e_{-}^{1},e_{-}^{3}\right] =2e_{-}^{3} & , & \left[ e_{-}^{2},e_{-}^{3}%
\right] =-e_{-}^{1} \\ 
&  &  &  &  \\ 
\left[ e_{+}^{1},e_{-}^{2}\right] =-2e_{+}^{2} & , & \left[
e_{+}^{1},e_{-}^{3}\right] =2e_{+}^{3} & , & \left[ e_{+}^{2},e_{-}^{1}%
\right] =2e_{+}^{2} \\ 
&  &  &  &  \\ 
\left[ e_{+}^{2},e_{-}^{3}\right] =-e_{+}^{1} & , & \left[
e_{+}^{3},e_{-}^{1}\right] =-2e_{+}^{3} & , & \left[ e_{+}^{3},e_{-}^{2}%
\right] =e_{+}^{1}%
\end{array}%
.
\end{equation*}

Next we introduce the map $\mathcal{G}:\mathfrak{h}_{+}\longrightarrow 
\mathfrak{h}_{-}$ defined as 
\begin{equation*}
\begin{array}{cc}
\mathcal{G}e_{+}^{i}=e_{-}^{i} & i=1,2,3%
\end{array}%
\end{equation*}%
which is $\mathfrak{h}_{-}$-invariant:%
\begin{equation*}
\mathcal{G}\left[ e_{-}^{i},e_{+}^{j}\right] =\left[ e_{-}^{i},\mathcal{G}%
e_{+}^{j}\right] ,
\end{equation*}%
and antisymmetric. Also, we introduce a linear operator $\mathcal{B}:%
\mathfrak{h}_{+}\longrightarrow \mathfrak{h}_{-}$ satisfying the $\mathcal{O}
$-operator condition $\left( \ref{Bop 02a}\right) $, realized through a
solution $\theta :\mathfrak{h}^{+}\longrightarrow \mathfrak{h}^{+}$ of the
modified classical Yang-Baxter equation on $\mathfrak{h}^{+}$ 
\begin{equation*}
\mathcal{B}e_{+}^{i}=\mathcal{B}\left( b_{i}+a_{i}\right) =\theta
b_{i}-\varphi \left( \theta \left( b_{i}\right) \right) .
\end{equation*}%
We use the quasitriangular factorizable solution of the modified classical
Yang-Baxter equation $2r\in \mathfrak{sl}_{2}\otimes \mathfrak{sl}_{2}$
introduced in eq. $\left( \ref{sl2 01}\right) $ and define $\theta \left(
b_{i}\right) =-\left( b_{i}\otimes I,r^{-}\right) $ from $r_{-}=b_{2}\otimes
b_{3}-b_{3}\otimes b_{2}$ so that

\begin{equation*}
\begin{array}{ccccc}
\theta \left( b_{1}\right) =0 & , & \theta \left( b_{2}\right) =b_{2} & , & 
\theta \left( b_{3}\right) =-b_{3}%
\end{array}%
.
\end{equation*}%
and the map $\mathcal{B}$ on the basis $\left\{
e_{+}^{1},e_{+}^{2},e_{+}^{3}\right\} \subset \mathfrak{h}_{+}$ is 
\begin{equation*}
\begin{array}{ccccc}
\mathcal{B}e_{+}^{1}=0 & , & \mathcal{B}e_{+}^{2}=e_{-}^{2} & , & \mathcal{B}%
e_{+}^{3}=-e_{-}^{3}%
\end{array}%
.
\end{equation*}%
From it we calculate the new Lie bracket on $\mathfrak{h}_{+}$, namely $%
\left[ X,Y\right] _{\mathcal{B}}=\left[ \mathcal{B}X,Y\right] -\left[ 
\mathcal{B}Y,X\right] $, which turns out to be%
\begin{equation}
\begin{array}{ccccc}
\left[ e_{+}^{1},e_{+}^{2}\right] _{\mathcal{B}}=-2e_{+}^{2} & , & \left[
e_{+}^{1},e_{+}^{3}\right] _{\mathcal{B}}=-2e_{+}^{3} & , & \left[
e_{+}^{2},e_{+}^{3}\right] _{\mathcal{B}}=0%
\end{array}
\label{BLbracket 01}
\end{equation}%
and we denote it as $\mathfrak{h}_{+}^{\mathcal{B}}$, which can be
identified as an $\mathfrak{sl}_{2}^{\ast }$ Lie algebra.

The next step is to construct the twilled action, so we need the bilinear
map $\sigma :\mathfrak{h}_{-}\longrightarrow \mathrm{End}\left( \mathfrak{h}%
_{+}\right) $ $\left( \ref{ybo 17f2}\right) $ and $\rho :\mathfrak{h}%
_{+}\longrightarrow \mathrm{End}\left( \mathfrak{h}_{-}\right) $ $\left( \ref%
{LbgB 00}\right) $, whose nonvanishing actions are%
\begin{equation*}
\begin{array}{lll}
\sigma _{e_{-}^{1}}e_{+}^{2}=-2e_{+}^{2} & , & \sigma
_{e_{-}^{1}}e_{+}^{3}=2e_{+}^{3} \\ 
&  &  \\ 
\sigma _{e_{-}^{2}}e_{+}^{1}=2e_{+}^{2} & , & \sigma
_{e_{-}^{2}}e_{+}^{3}=-e_{+}^{1} \\ 
&  &  \\ 
\sigma _{e_{-}^{3}}e_{+}^{1}=-2e_{+}^{3} & , & \sigma
_{e_{-}^{3}}e_{+}^{2}=e_{+}^{1}%
\end{array}%
\end{equation*}%
and 
\begin{equation*}
\begin{array}{ccc}
\rho \left( e_{+}^{1},e_{-}^{2}\right) =-2e_{-}^{2} & , & \rho \left(
e_{+}^{1},e_{-}^{3}\right) =-2e_{-}^{3}%
\end{array}%
.
\end{equation*}

The twilled extension $\mathfrak{g}_{\mathcal{B}}=\left( \mathfrak{h}_{+}^{%
\mathcal{B}}\right) ^{op}\oplus \mathfrak{h}_{-}$ of $\left( \mathfrak{h}%
_{+}^{\mathcal{B}}\right) ^{op}$ and $\mathfrak{h}_{-}$ is then defined by
the Lie bracket 
\begin{equation*}
\begin{array}{l}
\left[ X_{+}+X_{-},Y_{+}+Y_{-}\right] \\ 
=-\left[ X_{+},Y_{+}\right] _{\mathcal{B}}+\left[ X_{-},Y_{+}\right] -\left[
Y_{-},X_{+}\right] \\ 
~~~+\left[ X_{-},Y_{-}\right] -\left[ \mathcal{B}X_{+},Y_{-}\right] -%
\mathcal{B}\left[ Y_{-},X_{+}\right] ~~~+\mathcal{B}\left[ X_{-},Y_{+}\right]
+\left[ \mathcal{B}Y_{+},X_{-}\right]%
\end{array}%
,
\end{equation*}%
giving rise to the nonvanishing crossed Lie brackets%
\begin{equation*}
\begin{array}{lll}
\left[ e_{+}^{1},e_{-}^{2}\right] _{\mathcal{B}}=-2e_{+}^{2}-2e_{-}^{2} & ,
& \left[ e_{+}^{1},e_{-}^{3}\right] _{\mathcal{B}}=2e_{+}^{3}-2e_{-}^{3} \\ 
&  &  \\ 
\left[ e_{+}^{2},e_{-}^{1}\right] _{\mathcal{B}}=2e_{+}^{2} & , & \left[
e_{+}^{2},e_{-}^{3}\right] _{\mathcal{B}}=-e_{+}^{1}+e_{-}^{1} \\ 
&  &  \\ 
\left[ e_{+}^{3},e_{-}^{1}\right] _{\mathcal{B}}=-2e_{+}^{3} & , & \left[
e_{+}^{3},e_{-}^{2}\right] _{\mathcal{B}}=e_{+}^{1}+e_{-}^{1}%
\end{array}%
\end{equation*}%
that is completed with%
\begin{equation*}
\begin{array}{ccccc}
\left[ e_{-}^{1},e_{-}^{2}\right] =-2e_{-}^{2} & , & \left[
e_{-}^{1},e_{-}^{3}\right] =2e_{-}^{3} & , & \left[ e_{-}^{2},e_{-}^{3}%
\right] =-e_{-}^{1}%
\end{array}%
\end{equation*}%
and $\left( \ref{BLbracket 01}\right) $. It coincides with the crossed
brackets of the \emph{twilled} \emph{extension} of $\mathfrak{sl}_{2}$ and $%
\left( \mathfrak{sl}_{2}^{\ast }\right) ^{op}$ namely $\left( \mathfrak{sl}%
_{2}\oplus \left( \mathfrak{sl}_{2}^{\ast }\right) ^{op}\right) _{\mathcal{B}%
}$, obtained in subsection \ref{ex. sl2 1}, after the substitutions $%
e_{-}^{1}=H$, $e_{-}^{2}=X_{-}$, $e_{-}^{3}=X_{+}$, $e_{+}^{1}=4h$, $%
e_{+}^{2}=2x_{+}$ and $e_{+}^{3}=2x_{-}$.

The orthogonal subspaces%
\begin{equation*}
\mathrm{E}_{\mathcal{B}}^{\pm }=\overline{\left\{ e_{+}^{i}+\left( \mathcal{B%
}\pm \mathcal{G}\right) e_{+}^{i}/i=1,2,3\right\} }
\end{equation*}%
are spanned by the sets $\left\{
e_{+}^{1}+e_{-}^{1},e_{+}^{2}+2e_{-}^{2},e_{+}^{3}\right\} $ and $\left\{
e_{+}^{1}-e_{-}^{1},e_{+}^{2},e_{+}^{3}-2e_{-}^{3}\right\} $, and one may
easily verify that they are Lie ideals in $\left( \mathfrak{sl}_{2}\oplus
\left( \mathfrak{sl}_{2}^{\ast }\right) ^{op}\right) _{\mathcal{B}}.$

\section{Conclusions}

We have studied some algebraic aspects of quadratic vector spaces with
Lagrangian and orthogonal splittings associated with complex product
structures. This algebraic structure comes to complete the role of the
involutive operator $\mathcal{E}$ which is well known in various settings
such as string theory, Poisson Lie T-duality and generalized K\"{a}hler
geometry, by associating a complex structure $\mathcal{J}$ constructed from $%
\mathcal{E}$. In addition, this structure can be tied to a generalized
metric on one of the Lagrangian components, and this is the origin of the
physical interest on it, since it arise as twisting of a genuine metric.
When this metric is encoded in the operator $\mathcal{E}$, the duality idea
becomes more apparent because a metric in the Lagrangian complement, related
to the former, is introduced.

When dealing Lie bialgebras, we have shown that starting from a Manin triple 
$\left( \mathfrak{g},\mathfrak{g}_{+},\mathfrak{g}_{-}\right) $ with a
generalized metric $\mathcal{G}+\mathcal{B}:\mathfrak{g}_{+}\longrightarrow 
\mathfrak{g}_{-}$, that gives rise to an orthogonal splitting $\mathfrak{g}=%
\mathcal{E}^{+}\oplus \mathcal{E}^{-}$, we can build the Manin triple $%
\left( \mathfrak{g},\left( \mathfrak{g}_{+}^{\mathcal{B}}\right) ^{op},%
\mathfrak{g}_{-}\right) $ where $\mathcal{E}^{+}$ and $\mathcal{E}^{-}$
become anti-isomorphic Lie ideals provided $\mathcal{B}$ is an $\mathcal{O}$%
-operator with extension $\mathcal{G}$ of mass $-1$. This can be interpreted
as $\mathfrak{g}=\mathcal{E}^{+}\oplus \mathcal{E}^{-}$ is a decoupling of
the Manin triple because there are no non trivial dressing action.
Conversely, starting from a quadratic Lie algebra direct sum of a pair of
orthogonal Lie algebras $\mathrm{E}^{+}$ and $\mathrm{E}^{-}$ we built a
quasi-Manin triple $\left( \mathfrak{g},\mathrm{F}_{+},\mathrm{F}_{-}\right) 
$ with a metric $\mathcal{G}$ on the $\mathrm{F}_{-}$-module $\mathrm{F}_{+}$%
. Then, after introducing a twisting $\mathcal{B}:\mathrm{F}%
_{+}\longrightarrow \mathrm{F}_{-}$, a Lie algebra structure $\mathrm{F}%
_{+}^{\mathcal{B}}=\left( \mathrm{F}_{+},\left[ ,\right] _{\mathcal{B}%
}\right) $ is defined on the vector space $\mathrm{F}_{+}$ by taking $%
\mathcal{B}$ as an $\mathcal{O}$-operator with extension $\mathcal{G}$ of
mass $-1$ and, in turn, we get a Main triple $\left( \mathfrak{g}_{\mathcal{B%
}},\left( \mathrm{F}_{+}^{\mathcal{B}}\right) ^{op},\mathrm{F}_{-}\right) $
where the orthogonal subspaces $\mathrm{E}^{+}$ and $\mathrm{E}^{-}$ are Lie
ideals.

In both cases, the $\mathcal{O}$-operator is a masked version of a
quasitriangular factorizable solution of the classical Yang-Baxter equation
so the key behind these constructions is the fact that if $\mathfrak{k}$ is
a quasitriangular factorizable Lie bialgebra then its double $\mathfrak{k}%
\oplus \left( \mathfrak{k}^{\ast }\right) ^{op}$ admits an orthogonal
splitting in anti-isomorphic Lie ideals, as shown in subsection 4.7.
However, the approach through generalized metrics and $\mathcal{O}$%
-operators allows making a direct contact with some subjects with geometric
contents as those described at the Introduction and in the end of Section 1.

From the point of view of possible application in some Theoretical Physics
problems, it is interesting to see this procedure as a way to obtain the 
\emph{decoupled modes} inside a Manin triple, since there are no nontrivial
dressing actions in the orthogonal Lie ideals splitting.

\end{document}